\documentclass{he_symp}
\usepackage{psfig,graphicx,epsfig}
\usepackage{color}
\usepackage{amsmath,amssymb,epic,eepic,array}
\begin{document}
\renewcommand{\FirstPageOfPaper }{ 64}\renewcommand{\LastPageOfPaper }{ 86}

\newcommand{\rxx}{RX J0852.0-4622}
\newcommand{\cxx}{CXOU J085201.4-461753}
\newcommand{\lapr}{\raisebox{-.6ex}{\mbox{$\stackrel{<}{\mbox{\scriptsize$\sim$}}\:$}}}
\newcommand{\degree}{{\rm o}}
\setcounter{page}{63}

\title{X-ray Observations of Neutron Stars and Pulsars:\\[-1.5ex] \linebreak First Results from XMM-Newton}
\author{W.Becker, B.Aschenbach }  
\institute{Max--Planck--Institut f\"ur extraterrestrische Physik,
 Giessenbachstra{\ss}e, 85741 Garching, Germany}
\authorrunning{W.Becker \& B.Aschenbach}
\titlerunning{X-ray Observations of Neutron Stars and Pulsars:  First Results from XMM}

\maketitle

\begin{abstract}

 The X-ray Multi-Mirror Mission XMM-Newton is ESA's largest observatory so far; it is 
 dedicated to explore the Universe in the $0.2 - 15$ keV X-ray band of the 
 electromagnetic spectrum. Because of its large collecting area very faint sources 
 not accessible before can be observed and it is therefore the long awaited instrument to
 study young pulsars and neutron stars in supernova remnants, cooling neutron stars 
 and millisecond pulsars at X-ray energies. The high throughput of the instruments,  
 which all are operated simultaneously, provide high resolution spectral, spatial 
 and temporal information from a source during a single observation and make XMM-Newton 
 unique and best suited for pulsar studies.
 In this article we briefly describe the instrument capabilities useful for
 pulsar observations and provide information on the timing accuracy on the 
 relative and absolute scale. We further provide scientific results from observations 
 of the Crab-pulsar, PSR J1617-5055 near RCW 103, of young neutron stars in 
 the supernova remnants RX J0852-4622, Puppis-A and RCW 103 including 
 1E161348-5055.1 which is identified  to be the second binary in a supernova remnant.
 In addition we report on observations of the cooling neutron star PSR B1055-52 
 and on the millisecond pulsar PSR J0030+0451 which all were observed by XMM-Newton 
 during the first two years of scientific operation.
\end{abstract}

\section{Introduction} \label{introduction}

 X-ray observatories like ROSAT, ASCA, BeppoSAX and RXTE have achieved 
 an important progress in neutron star and pulsar astronomy. The 
 identification of Geminga and of cooling neutron stars (Halpern \& Holt 
 1992; \"Ogelman 1995 and references therein), the discovery 
 of pulsed X-ray emission from millisecond pulsars (Becker \& Tr\"umper 
 1993) and of the first accreting X-ray ms-pulsar SAX J1808.4-3658 
 (in't Zand et al.~1998; Wijnands \& van der Klis) are only a few of 
 the fascinating results.
 ROSAT's predominant contribution was due to its favorable soft response down to 
 0.1 keV, which permitted to observe the very soft thermal emission from the 
 neutron stars' surface, but the relatively small bandwidth of up to 2.5 keV 
 precluded any broad band study of the pulsar emission. Any conclusive 
 identification of the nature of the 'hard tails' indicated in various X-ray spectra 
 of cooling neutron stars and ms-pulsars was therefore not possible. 
 ASCA and the Narrow Field Instruments on BeppoSAX, on the other hand, 
 were sensitive up to 10 keV and were very successful in discovering 
 young neutron stars in SNRs by their hard X-ray emission (cf.~Torii 
 1998; Saito 1998). The lack of high resolution imaging, however, did 
 in most cases not allow to resolve any central compact X-ray  
 source from the surrounding plerion or supernova remnant.

 The expectations from an observatory like XMM-Newton, with its 
 enormous collecting area, multi-instrumentation and wide energy 
 bandwidth are therefore high, especially in the field of neutron 
 star astronomy where the targets (apart from a few Crab-like objects)
 are known to be relatively faint X-ray emitters.  

 XMM carries three identical X-ray telescopes, each of which is 
 equipped with an X-ray CCD camera in the focal planes, called  
 EuroPean Imaging Camera (EPIC). Two of the cameras are constructed 
 using Metal Oxide Semiconductor technology (MOS1 \& MOS2) and the third 
 camera is built in Positive-Negative depleted
 Silicon Semiconductor technology, therefore called PN-camera. 
 About half of each of the beams of the two telescopes operating with the MOS's 
 is intercepted by Reflection Grating Spectrometers (RGS1 \& RGS2), 
 which diffract the X-rays to two dedicated CCD detectors. Optical 
 observations can be made with a separate co-aligned optical telescope 
 called Optical Monitor (OM). The instruments can be operated   
 in parallel and they provide various observation 
 modes which are very well suited for pulsar studies. Table \ref{EPIC_Modes} 
 summarizes the timing modes available for the PN and MOS detectors. 
 Also listed are the detectors' field of view (in units of detector pixels and 
 angular units) and the associated temporal resolution. Useful  
 is the PN small-window mode which provides imaging information for
 a sky field of $\sim 4.4' \times 4.4'$ along with spectral and
 a $\sim 6$ ms temporal resolution. This mode has its benefit  
 for timing studies of sources located in crowded regions or in 
 supernova remnants. It prevents source confusion which often is 
 found to be the limiting factor when using the faster timing modes. 
 For the PN fast-timing and burst modes the spatial and spectral 
 information from a $64 \times 199$ CCD pixel array is condensed 
 into a one dimensional 
 $64 \times 1$ pixel array (1D-image), i.e. the spatial information in 
 Y-direction is lost due to the continuous read-out of the CCD. 
 In these modes, the complete photon flux (source plus DC emission from
 foreground or background sources located along the  read-out 
 direction) is accumulated and collapsed in the final 1D-image, 
 severely reducing the signal-to-noise ratio and preventing  
 the detection of X-ray pulsations from the target of interest.
 Soft-proton flares which are known to happen on various  
 time scales and frequencies almost unpredictable  can likewise produce enough 
 DC flux to prevent the detection of X-ray pulsations from a 
 pulsar of intermediate X-ray brightness if the fast-timing 
 mode was selected for the observation. For pulsars with 
 periods longer than $\sim 60$ ms, the PN small-window mode 
 is a good choice even in view of a $\sim 30\%$ reduced 
 efficiency compared with the fast-timing mode. For millisecond
 pulsar observations, however, the fast-timing mode is the only 
 one which provides sufficient temporal resolution for timing 
 studies.

\begin{table}
\caption{Instrument modes available  on
 XMM-Newton for timing studies of pulsars. 
 The image size is given in detector pixels and angular units.
 FOV is the full field of view.} \label{EPIC_Modes}
\begin{small}
\begin{tabular}{l l r}
 EPIC PN       &   \quad \quad$\!\!\!\!\!\!$ Image  size / \quad $\,\,$ FOV  & Temporal\\
  Mode         &       \quad \quad $\,\,$ pixel \quad / \quad arcmin    &  resolution\\\hline\\[-2ex]
 Small window  &  2D - $64 \times \,\,64$  / $4.4' \times 4.4'$ &    5.67 ms \\
 Fast-timing   &  1D - $64 \times 199$     / $4.4' \times 13.75'$    &  0.02956 ms\\
 Burst         &  1D - $64 \times 199$     / $4.4' \times 13.75'$    &  0.0072 ms\\
\end{tabular}\\[3ex]

\begin{tabular}{l l r}
 EPIC MOS       &   \quad \quad$\!\!\!\!\!\!$ Image  size / \quad $\,\,$ FOV  & Temporal\\
  Mode          &       \quad \quad\quad pixel \quad / \quad arcmin    &  resolution\\\hline\\[-2ex]
 part.~window1  &  2D - 100 x 100 $\!$ / 1.8' x 1.8'  &     200 ms \\
 part.~window2  &  2D - 300 x 300 $\!$ / 5.5' x 5.5'  &     700 ms \\
 timing         &  1D - user defined pixel area       &   $\ge$ 1.5 ms\\
\end{tabular}
\end{small}\\

 \begin{center}
 \begin{minipage}{8.5cm}\footnotesize
  Note: In PN timing modes only CCD \#4 (of 12 CCDs in total) is active whereas in MOS timing
  modes only the central CCD provides timing information and the remaining 6 chips are operated
  in imaging mode. Burst mode is designed for very bright sources up to 6.3 Crab only. The efficiency
  of the PN small-window mode is  $\sim 30\%$ of that of the PN timing mode. In burst mode it is
  only 3\%. XMM's spatial resolution  is $\sim 15"$ (HEW). The size of a CCD pixel is $1.1" \times 1.1"$
  and $4.125" \times 4.125"$ for the MOS and PN camera, respectively.
 \end{minipage}
 \end{center}
 \end{table}

 The strong interest in XMM-Newton, especially for neutron star and
 pulsar research, is supported by the large list of sources accepted
 for observations in AO1 (cf.~Table \ref{ao1_targets}). The proposed
 science and accepted targets cover a wide range of different categories,
 from young pulsars in supernova remnants to cooling neutron
 stars, binary-pulsars, old but nearby pulsars and millisecond pulsars.
 Even a small survey for potentially X-ray bright rotation-powered pulsars 
 is included.

 In the following sections we will report on the first XMM observations
 of pulsars, browsing through the various categories from young Crab-like
 pulsars and neutron star candidates in supernova remnants to the very old
 millisecond pulsars, summarizing the most exciting results. Before we
 do so, however, we will briefly review the various emission processes
 discussed to be the source for the observed X-ray emission.

\section{High-energy X-ray Emission Processes} \label{photospheric}

  As a result of observations with the satellite observatories ROSAT, ASCA and
  Chandra, the number of rotation-powered  pulsars seen at X-ray
  energies has reached 51 (as of June 2002). Almost half of the
  detected pulsars are ms-pulsars. Several of these objects have been detected at 
  optical wavelengths and at gamma-ray energies (cf.~Tab.\ref{psr_sym_tab} and 
  Tab.\ref{ms_psr_sym_tab} for a summary), advocating for multi-wavelength studies 
  of the pulsar emission.  This approach is of great benefit as the physical 
  processes which cause the emission in different wavelength bands are 
  obviously related to each other. Although the quality of the data obtained 
  with different instruments and/or in different wave-bands is inevitably 
  rather inhomogeneous,  and the conclusions drawn on these data are to 
  some extent unlikely to be solid in a number of cases, there is general consensus that the 
  X-radiation detected from rotation-driven pulsars has to be attributed 
  to various thermal and non-thermal emission processes including the following:

 \begin{itemize}

  \item[$\bullet$]  Non-thermal emission from charged relativistic particles
         accelerated in the pulsar magnetosphere. This emission is
         characterized by power-law like spectra over a  broad energy
         band. The emitted radiation can be observed from the
         optical to the gamma-ray band.\\[-2ex]

  \item[$\bullet$]  Extended emission from pulsar-driven synchrotron nebulae.
         Depending on the local conditions (e.g., matter density of the ambient
         medium), these nebulae can be observed from
         radio through hard X-ray energies.\\[-2ex]

  \item[$\bullet$]  X-ray and gamma-ray emission from interaction of relativistic
         pulsar winds with a close companion star or with the wind of
         a companion star.\\[-2ex]

  \item[$\bullet$]  Photospheric emission from the hot surface of a cooling neutron
         star. In this case a modified blackbody spectrum and smooth,
         low-amplitude intensity variations with the rotational period
         are expected, observable from the optical through the soft X-ray
         range.\\[-2ex]

\end{itemize}

 \begin{table*}
  \caption{List of Pulsars which are accepted for observations by XMM-Newton during AO1. Also
  listed is the main instrument (PN, MOS, RGS or OM) and the proposed exposure time in ksec.
  CAL indicates calibration targets which are observed at regular time intervals for calibration
  purposes and in various instrument modes and setups.  {\em obs} indicates whether a target
  has been observed by June 2002. Category C-Targets are supposed to be optional and not
  guaranteed to be observed.} \label{ao1_targets}
  \begin{picture}(172,130)(0,3)
  \put(0,0){\centerline{\psfig{figure=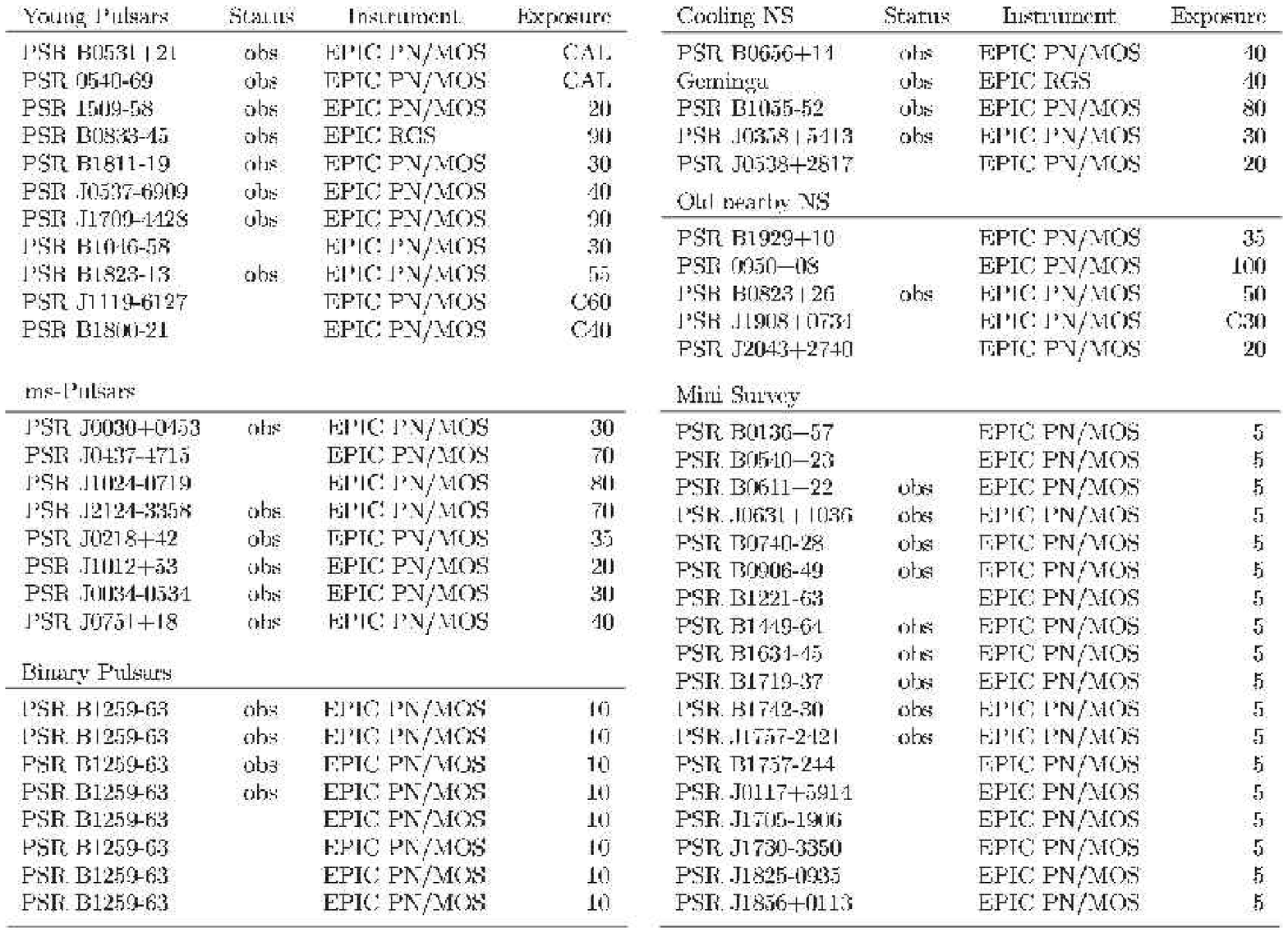,width=18cm,clip=}}}
  \end{picture}
 \end{table*}

 In many pulsars the observed X-ray emission is due to a mixture of
 different thermal and non-thermal processes. The available data do not
 always allow to discriminate  between the different emission scenarios.
 This was true for ROSAT and ASCA observations and might still be true for Chandra
 and XMM data. 

 It is well established, however, that magnetospheric
 emission from charged particles, accelerated in the neutron star magnetosphere
 along the curved magnetic field lines, dominates the radiation from young
 {\em rotation-powered} pulsars with ages $\lapr 5000$ years. Middle aged pulsars 
 such as the three Musketeers PSR B0656+14, PSR B1055-52 and Geminga are seen to 
 have composite spectra consisting of two or three components which are a soft 
 thermal (cooling) and a hotter polar-cap component as well as some hard emission 
 associated with non-thermal processes (cf.~\S\ref{cooling_ns}). 
 For the millisecond pulsars, ROSAT and ASCA did not allow to conclusively
 clarify whether their emission is purely due to the presence of hot/heated 
 polar-caps, due to non-thermal magnetospheric emission or a mixture of both
 or whether there exists even a dichotomy among the emission properties 
 which correlates with the pulsar's spin-parameters. Chandra and XMM-Newton
 observations of the two ms-pulsars PSR J0437-4715 and PSR J0030+0451 
 (cf.~\S\ref{ms_psrs}) as well as the detection of 11 ms-pulsars in 
 the Globular Cluster 47TUC seem to suggest that both, a thermal and a 
 non-thermal emission component are present in all ms-pulsars though it
 is difficult to accomplish the gross similarity between the radio and 
 X-ray pulse profiles with these two (or even more) different 
 components at work.

\section{Young Neutron Stars in Supernova Remnants}

\subsection{The Crab- and Vela-like Pulsars} \label{crab_like_ns}

  The majority of the young Crab- and Vela-like pulsars have been observed by 
  XMM-Newton during AO1 with the aim to study the pulsars' temporal emission 
  properties along with a study of the pulsars' wind nebulae or a search for it. 
  While a large fraction of data was delivered to the PIs only recently and are 
  still in the process of data analysis, the Crab, PSR 0540-69 and PSR 1509-58 
  were observed for the purpose of instrument calibration (clock and mirrors),
  and have undergone a first scientific analysis already.

\begin{figure}[t!!!!]
   \centerline{\psfig{file=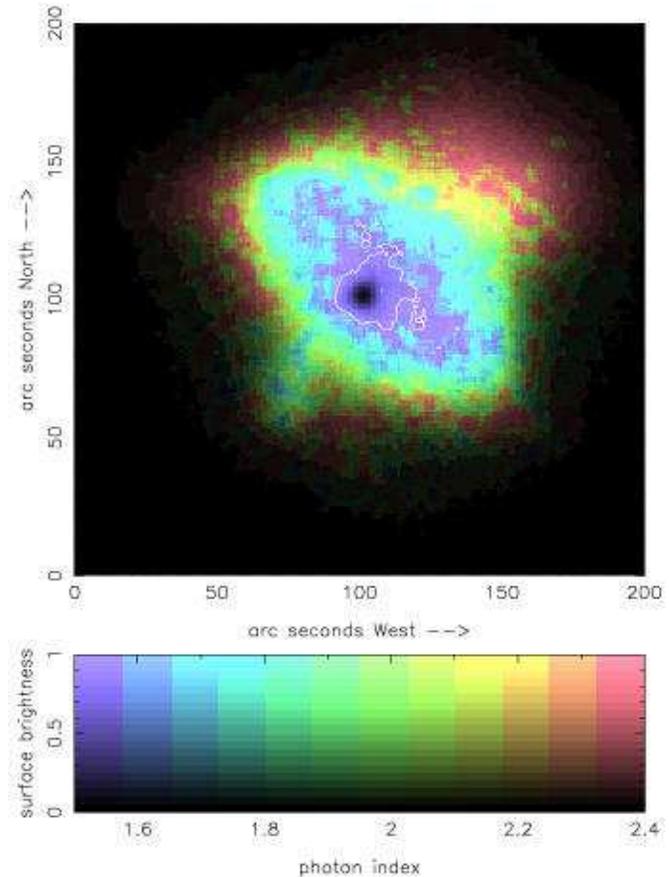,width=8.8cm,height=11.8cm,clip=}}
   \caption[]{\renewcommand{\baselinestretch}{1.0}\small\normalsize\small
    The Crab as observed with the European Imaging Camera
    aboard XMM-Newton. The image has been spatially oversampled, using a bin size
    of 1 arcsec (instrument HEW $\sim 15$ arcsec). The color coding of surface
    brightness and photon index is shown in the lower panel. The white contour
    encircles the region compromised by pile-up. (From Willingale et al.~2001)
    \label{xmm_crab_photon_index}} 
 \end{figure}

  Although the Crab has been observed with almost every instrument since its 
  discovery, it is amazing to see how many new results were obtained
  with Chandra and XMM-Newton. First images taken with the Advanced
  CCD Imaging Spectrometer (ACIS) aboard Chandra have already provided
  spectacular details of the inner nebula structure associated with the
  pulsar-wind outflow --- in addition to the torus ($r\approx 0.38$ pc)
  the {\em inner ring} ($r\approx 0.14$ pc), {\em jet} and {\em counter-jet}
  have been identified (Weisskopf et al.~2000). In the same observation,
  systematic variations of the X-ray spectrum were discovered  throughout
  the nebula. A more detailed study of the spectral variations became
  possible with the XMM data, taken in March 2000 with the EPIC detectors
  (Willingale et al.~2001). Figure \ref{xmm_crab_photon_index} shows an
  X-ray image of the Crab nebula in which the color coding is associated
  with the photon-index $\alpha$ of the nebula's X-ray spectrum, $dN/dE
  \propto E^{-\alpha}$. This image impressively illustrates the
  different shape of the spectra of the torus ($\alpha=1.8\pm 0.006$),
  the jet ($\alpha=2.1\pm 0.013$) and the outer nebula regions ($\alpha=
  2.34\pm 0.006$). Similar results were obtained by Chandra, measuring
  the hardness ratio distribution throughout the nebula (Weisskopf et
  al.~2000). For the pulsar, a photon spectral index of $\alpha=1.63 \pm 0.09$ is
  determined from out-of-time events selected along the read-out direction
  of the CCDs, and thus not affected by pile-up. The spectral difference 
  between the jet and the torus is found to be likely due to an intrinsically 
  steeper electron spectrum of the jet. The outer regions of the nebula show 
  the steepest spectrum, which is likely to be due to enhanced synchrotron losses 
  of the electrons during their ride from the pulsar to the outskirts 
  (Willingale et al.~2001).

\begin{figure}[t]
   \centerline{\psfig{file=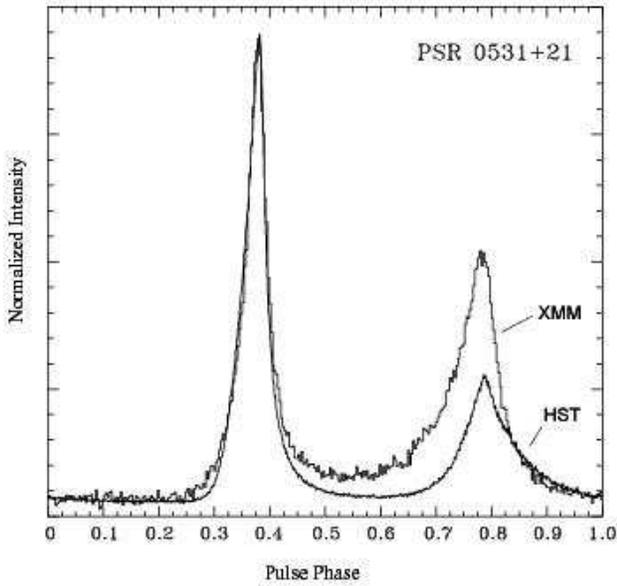,width=8.2cm,clip=}}
   \caption[]{\renewcommand{\baselinestretch}{1.0}\small\normalsize\small The Crab pulse profile
    observed with the XMM-Newton and Hubble Space observatories. Peaks of the main pulse have been 
    co-aligned in phase. A slight phase difference between the main and the second peaks is 
    apparent which suggests a small energy dependence of the phase shift between main and 
    secondary pulse peak. \label{xmm_hst_crab_pulse_profile}}
 \end{figure}

 Observations with the High Resolution Imager (HRI) aboard ROSAT has already shown 
 that at least $\sim 75\%$ of the Crab-pulsar's total soft X-ray flux is emitted 
 from the co-rotating magnetosphere (Becker \& Tr\"umper 1997). Recent observations 
 with the High Resolution Camera aboard Chandra by Tennant et al.~(2001) at 
 an angular resolution of $\le 1$ arcsec has improved this number in showing that 
 in the $0.5-10$ keV band the pulsed fraction is $\sim 85\%$. The high pulsed 
 fraction along with the monthly published radio ephemeris (Lyne et al., 2002) 
 makes the Crab therefore an excellent target for the XMM-clock calibration. 
 The Crab-pulsar was observed for that reason in March 2000, 2001 and 2002 with 
 the PN in timing and burst mode, respectively. Because of the pulsar's brightness 
 only the burst mode provides accurate spectral information not affected by pile-up 
 effects, though the detector life time in that mode is only 3\%. We have analyzed 
 the Crab data from all orbits (cf.~Fig.\ref{xmm_hst_crab_pulse_profile}) and found 
 that the pulsar period deduced from the XMM data matches to better than 
 $2.5\times 10^{-10}\,\mbox{s}$ the radio period given in the Jodrell Bank 
 monthly Crab ephemeris (Lyne et al., 2002). The difference is within the expected 
 statistical uncertainty\footnote{In previous versions of XMM-SAS the barycenter 
 correction code was erroneous, leading to discrepancies of periods derived from 
 EPIC data and compared to, e.g.~radio ephemeris of up to $\Delta P/P = 7 \times 
 10^{-6}$ (Kuster et al.~2002). The code was fixed recently (SAS 5.3 and later 
 versions) which improved the EPIC relative timing by three orders of magnitude.}.

 In addition to the relative timing we checked the accuracy of the XMM clock on 
 the absolute scale by comparing the arrival time of the Crab pulsar's
 main peak with that measured in the radio band (Lyne et al., 2002). We find 
 that for all XMM data the arrival of the main peak lead the arrival of the 
 main radio peak by $0.047 \pm 0.001$ in phase (modulo the pulsar period), 
 corresponding to a time difference of 1.57 ms (modulo the pulsar period). 
 Rots et al.~(2000) have analyzed all Crab data taken with RXTE since the 
 start of the mission in January 1996 and found that the phase of the main  
 X-ray peak is different from the radio peak by $\sim 300 \mu$s. Depending 
 on whether this phase difference observed in RXTE data is real or only  
 reflects the accuracy of the RXTE timing, the phase alignment between 
 radio and X-ray pulses measured by XMM is uncertain by $1.27 -1.57$ ms 
 on the absolute scale. Careful analysis of all modules involved in XMM's timing 
 are currently in progress in order to improve the absolute clock accuracy.

   Besides the Crab several other young pulsars have been observed by XMM during the first
   two years of scientific observations. One of these pulsars is PSR J$1607-5055$, which is
   a 69 ms pulsar located about 7 arcmin outside the boundary of the young supernova remnant
   RCW 103 (cf.~Fig.\ref{xmm_rcw_103}).  The spin-down age of the pulsar is $\tau \sim 8000$
   years (cf.~Table \ref{psr_sym_tab}) placing it among the youngest known radio pulsars. An
   association of RCW 103 and the pulsar was discussed but found to be unlikely (Dickel \&
   Carter 1998; Kaspi et al.~1998). The pulsar distance is not very well constrained. The radio
   dispersion measure yields a distance of $6.1-6.9$ kpc (Taylor \& Cordes 1993) whereas Kaspi
   et al.~(1998) conclude from a comparison between dispersion measure based distances and
   distances obtained from HI absorption measurements of PSR 1641-45 and PSR 1718-35 - which are
   both located within $20^\degree$ of PSR J1607$-$5055 - that the pulsar could be as close
   as $\sim 4.5$ kpc.

\begin{figure}[t]
   \centerline{\psfig{file=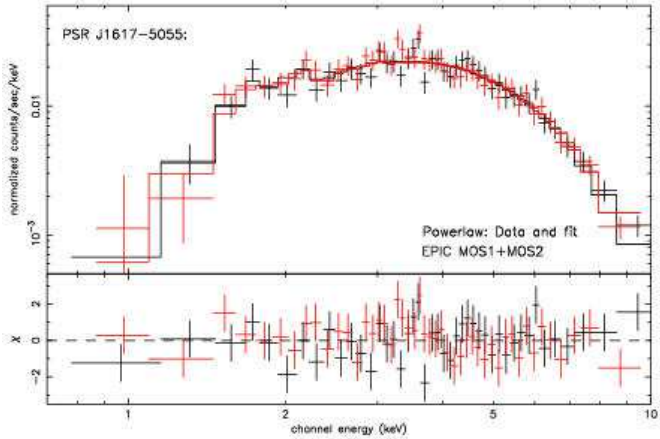,width=8.8cm,clip=} }
   \caption[]{\renewcommand{\baselinestretch}{1.0}\small\normalsize\small
    X-ray spectrum of PSR J1617-5055  taken with MOS1/2 and fit by a power-law.
    \label{PSR_J1607_spectrum} }
   \end{figure}

    PSR J$1607-5055$ was discovered  by its X-ray pulses in archival GINGA data (Torii et al.~1998).
   It was then detected also in the radio domain (Kaspi et al.~1998). With ASCA it was not
   possible to separate the pulsar from the close by RCW 103 supernova remnant, which limited
   modeling of the pulsar spectrum to energies $>$ 3 keV (Torii et al.~2000). XMM-Newton has
   observed the pulsar in September 2001 for 30 ksec and for another 20 ksec but off-center
   because the  target was RCW 103 (cf.~\S\ref{radio_silent_ns}). The XMM data of
   PSR J1617$-$5055 provide the first detailed pulsar spectrum in the $0.3-10$ keV band.
   Among various spectral models a power-law spectrum with a photon index of $\alpha=1.1-1.4$
   (90\% confidence range) is found to be the simplest but acceptable  description of the
   spectrum up to 10 keV ($\chi^2_\nu=0.93$ for 121 dof). Thus the X-ray emission is classified
   as non-thermal. The emission turns out to be highly absorbed, with $N_H = (2.8 - 3.6)
   \times 10^{22}\,\mbox{cm}^{-2}$, prohibiting any measurement of the softer cooling emission
   from the neutron star surface. Figure \ref{PSR_J1607_spectrum} shows the energy spectrum
   as observed by the EPIC MOS1/2. The X-ray flux in the $0.5-10$ keV band is $f_x= (4.9 - 5.4)
   \times 10^{-12}\,\mbox{erg s}^{-1}\mbox{cm}^{-2}$, implying a luminosity of $L_x= (2.1 - 2.3)
   \times 10^{34}\,\mbox{erg s}^{-1}$ for a distance of $d=6$ kpc. The inferred X-ray conversion
   efficiency is $L_x/\dot{E}= 1.4\times 10^{-3}$ and $3.6 \times 10^{-4}$ in the $0.5-10$ keV
   and $0.1 - 2.4$ keV band, respectively.

   Diffuse X-ray emission associated with PSR J1617$-$5055, either because of the presence of
   a plerion or a supernova remnant, has not been detected in the XMM-Newton data
   (cf.~Fig.\ref{xmm_rcw_103}). Chandra observed PSR J1617$-$5055 in September 1999 and
   February 2000.  We analyzed the public archive data, and found no indication of diffuse
   extended emission surrounding PSR J1617$-$5055 either (Becker et al., 2002a).

\begin{figure}[t!!!!!!!!!!]
   \centerline{\psfig{file=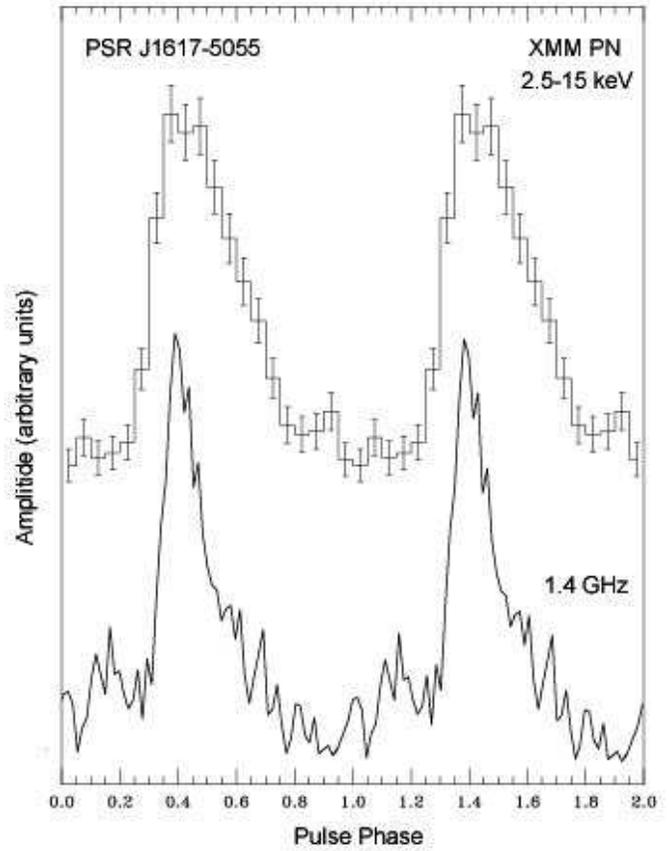,width=8.8cm,clip=} }
   \caption[]{\renewcommand{\baselinestretch}{1.0}\small\normalsize\small
    X-ray and radio pulse profiles of PSR J1617-5055 as observed in the $2.5 - 15$ keV band with the XMM 
    EPIC-PN and at 1.4 GHz with the Parkes Radio Telescope.  Both the X-ray and the radio profile show a single peak.
    The X-ray pulse has a duty cycle of $\sim 50\%$ in the 2.5-15 keV band.Two pulsation cycles are shown for clarity.
   (The radio pulse profile has been adopted from Kaspi et al.~1998) \label{J1617_lc} }
   \end{figure}

   In order to investigate the pulsar's temporal X-ray emission properties we used the data from
   the EPIC-PN camera which was operated in timing mode with a temporal resolution of
   0.03 ms (cf.~Tab.\ref{EPIC_Modes}). The presence of the nearby RCW 103 supernova remnant
   required a tight roll-angle constraint of $268^\degree -270^\degree$ for the lower and upper
   position angles in order to  minimize source confusion with photons being recorded along the
   read-out direction of the CCD columns in which PSR J$1617-5055$ was placed. To extract
   the pulsar events we selected an area of $8 \times 199$ pixels from the central CCD \#4
   of the PN camera. According to the fractional encircled energy function this means that we sample
   about 70\% of the source photons in $x$- and 100\% in $y$-direction. After
   barycentrisation (we used the source position  RA=16:17:29.45, DEC=-50:55:13.61) and
   correcting for the satellite orbit we searched for the pulsar period using an
   fft-based algorithm. To optimize the pulsar signal we blocked off all emission below
   2.5 keV as the  DC level at these energies is found to exceed the pulsar signal due to
   the emission from the nearby RCW 103, which reduced the signal-to-noise ratio and
   the detectability of the X-ray pulses significantly. But still with the remaining events we detected strong pulsed
   emission up to 15 keV. The pulse profile shows a single peak with a duty cycle of $\sim 50\%$
   and it shows a pronounced similarity with the radio profile at 1.4 GHz reported by Kaspi et
   al.~(1998). This suggests a close correlation between the radio and X-ray emission mechanisms
   (cf.~Fig.\ref{J1617_lc}). Comparing the X-ray pulse profile of PSR J$1617-5055$ with those
   observed from other young pulsars we find a striking similarity with PSR 1509-58 and  PSR 0540-69.
   For the epoch MJD=52155.5992851784, which is the mean epoch of the XMM observation in TDB at the
   solar system barycenter, we measure a period of $P=69.37252615720$ ms. The accuracy
   of the XMM clock against UTC is good enough to measure the arrival time of the X-ray
   pulse which we find at phase 0.65 (center of mass of the pulse) for the epoch given above.

  In order to investigate the  energy dependence of the X-ray pulse we performed a
   timing analysis separately for the $2.5-4.5$ keV, $4.5-9.0$ keV and $9.0-15$ keV bands. Neither the
   overall shape of the pulse nor its phase is found to change with energy. The pulsed
   fraction we measured in the different  energy bands using a  bootstrap method (Swanepoel
   et al.~1996) is $48 \pm 4 \%$, $57 \pm 4 \%$ and $69 \pm 19 \%$ in the $2.5-4.5$ keV,
   $4.5-9.0$ keV and $9.0-15$ keV bands, respectively. The fraction of the pulsed
   flux in the $2.5 - 15$ keV band is $53 \pm 3 \%$.

\begin{table*}[ht!!!!!!!!!!!!!!]
\begin{picture}(130,230)(13,45)
\put(0,0){\psfig{figure=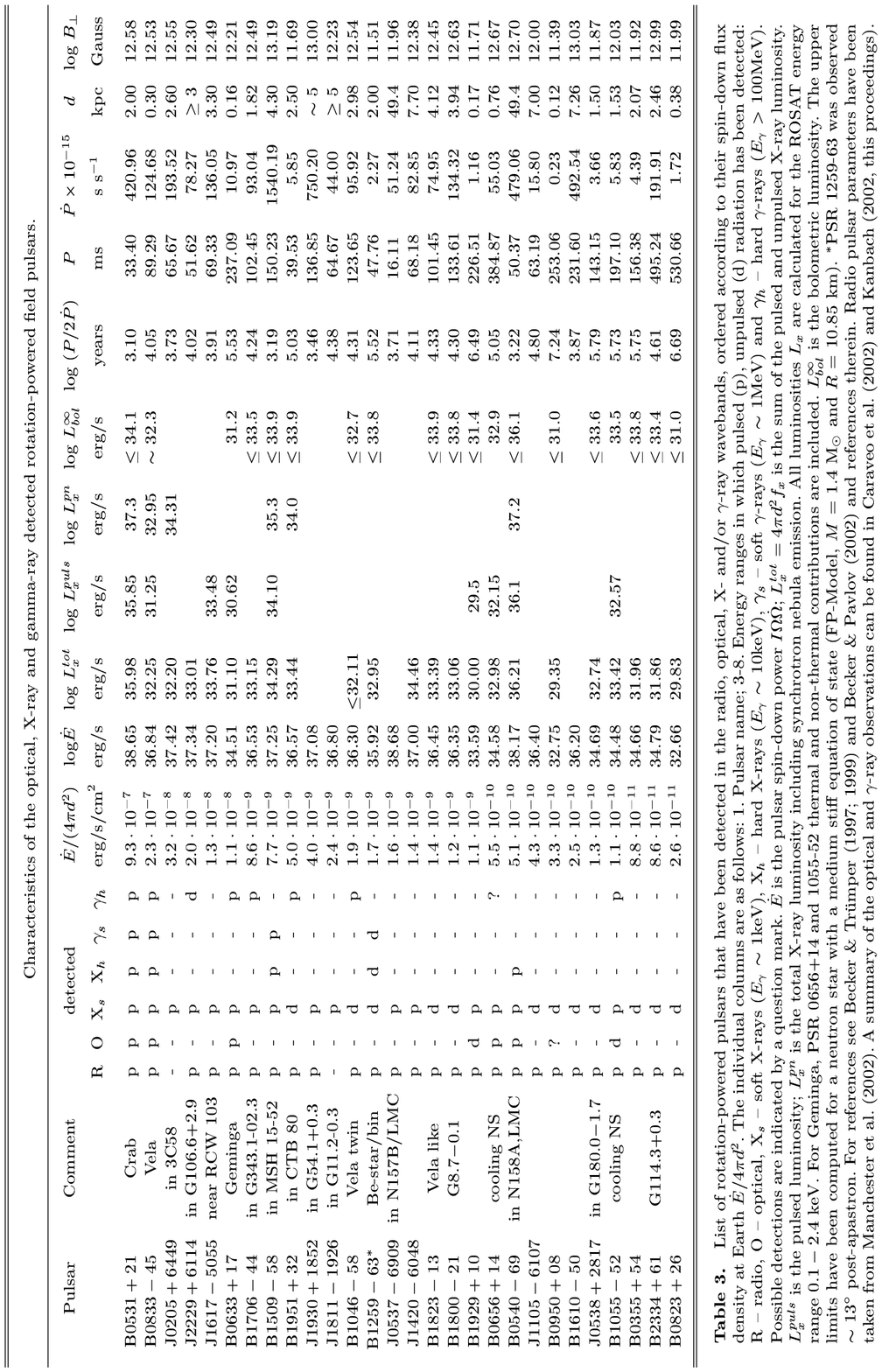,clip=}}
\end{picture}
\refstepcounter{table} \label{psr_sym_tab}
\end{table*}

\subsection{Radio-silent Neutron Stars in Supernova Remnants}  \label{radio_silent_ns}

\begin{center}
 \begin{figure}[t!!!]
  \centerline{\psfig{file=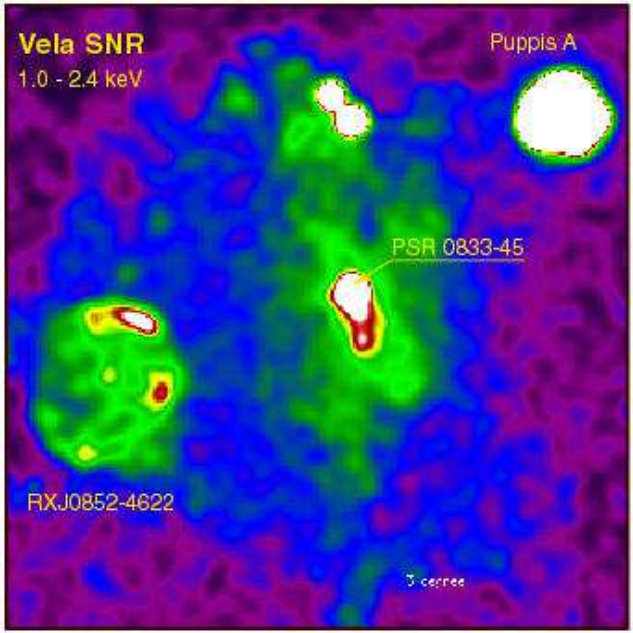,width=8.45cm,}}
  \caption[]{\renewcommand{\baselinestretch}{1.0}\small\normalsize\small
    Vela and friends: The two supernova remnants Puppis-A and \rxx\ are
    both located at the edge of the Vela supernova remnant. Emission from the 
    neutron star \cxx\ is clearly visible in the ROSAT all-sky survey image 
    near to the center of \rxx.} \label{Vela_Jr_Rosat}
  \end{figure}
  \end{center}

 A fascinating result reflecting the advantage of wide-field spectro-imaging
 observations provided by ROSAT in its all-sky survey (RASS) is the discovery
 of the supernova remnant \rxx\ by  Aschenbach (1998). The remnant is located  
 along the line of sight towards the south-east corner of the Vela supernova  
 remnant (cf.~Fig.\ref{Vela_Jr_Rosat}). Because the Vela SNR dominates the emission 
 in the soft band, \rxx\ is best visible in the RASS data at energies above 
 $\sim 1$ keV. The remnant has a circular shell-like shape with a diameter of 
 2 degrees. Although the distance to the remnant is quite uncertain, its 
 free-expansion age ($t\sim 3.4 \times 10^3 v_{5k}^{-1} d_{1kpc}\,\mbox{yr}$, 
 with $v_{5k}$ as expansion velocity in units of 5000 km/s) suggests a remnant
 age much younger than that of Vela.
 \rxx, often denoted as {\em Vela-Junior} because of its young age, turns out to be
 remarkable in several aspects. COMPTEL data suggest that the remnant has a 1.157 MeV
 $\gamma$-line from the radioactive isotope ${}^{44}$Ti exclusively produced in supernovae 
 (Iyudin et al.~1998). If true, the remnant could be as young as $\sim 620$ years and as close
 as $\sim 200$ pc (Aschenbach et al.~1999), making it the closest and youngest supernova
 remnant among the $\sim 250$ remnants known today (Green 2000).
 However, as the detection of the ${}^{44}$Ti line is of low significance, it leaves
 the age and distance estimates of the remnant widely unconstrained, but see Aschenbach 
 et al. (1999) for a detailed discussion. ASCA observations have shown that the X-ray 
 emission of the brightest parts of the remnant are dominated by non-thermal emission 
 (Tsunemi et al.~2000; Slane et al.~2001) with a column density about one order of 
 magnitude larger than what is known for the Vela supernova remnant. Together with SN1006 
 and G347.3-0.5, \rxx\ thus form the first members of  the exclusive group of non-thermal
 shell-type supernova remnants,  believed to be accelerators of cosmic rays.
 The lack of strong variation in $N_H$ across \rxx\ further indicates that the remnant
 cannot be more distant than the Vela Molecular Ridge (Slane et al.~2001), locating it
 most likely behind the Vela remnant at a distance between 1-2 kpc.\\[-8ex]

 The remnant and its emission properties are interesting by its own and justify
 deep and detailed X-ray studies, but a compact central X-ray source has been detected in the
 remnant as well. First recognized as a $5\sigma$ excess in the RASS data by Aschenbach et
 al.~(1998) its existence was supported by the detection of a hard, somewhat
 extended X-ray source in the ASCA data (Tsunemi et al.~2000; Slane et al.~2001). To
 conclusively identify this source as the compact remnant of the supernova explosion
 that created \rxx\ was not possible because of the presence of two foreground
 stars (HD 76060 and Wray 16-30) which could be responsible for at least part of the
 central X-ray emission. But a recent observation with the Chandra ACIS-I has clearly
 invalidated the association with either one of these stars (Pavlov et al.~2001) and 
 another faint but hard X-ray source found by Mereghetti (2001) in a recent BSAX 
 observation. In view of this, there is little doubt that the central X-ray source
 \cxx, located only 4 arcmin off the geometrical center of \rxx,  indeed is the compact
 remnant of the supernova explosion. This is further supported by the $f_x/f_{opt}$
 ratio as the source has no optical counterpart brighter than $m_B \sim 22.5$.
 The short 3 ksec Chandra observation provides only an accurate position for \cxx. 
 Spectral analysis is strongly hampered by small photon statistics and pile-up 
 effects (Pavlov et al.~2001). There is no extended X-ray emission reported from 
 this short Chandra observation.

\begin{figure}[t!!!]
 \psfig{file=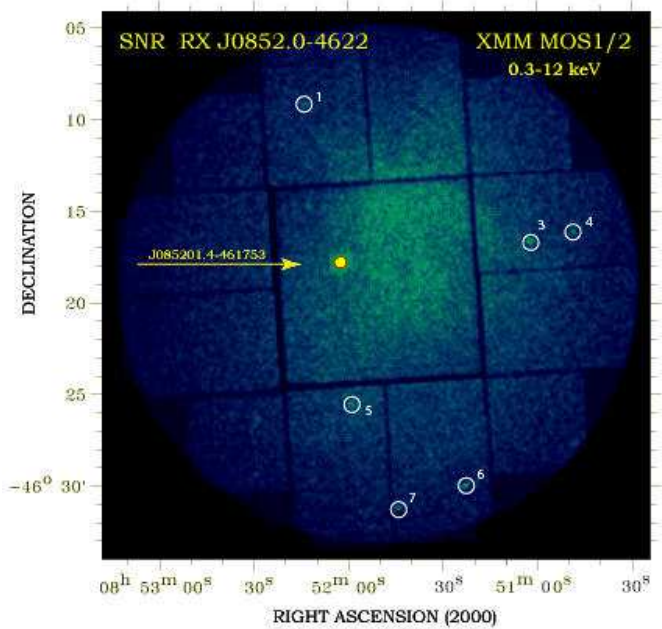,width=8.8cm}
 \caption[]{\renewcommand{\baselinestretch}{1.0}\small\normalsize\small
  XMM's view of the $\sim 30$ arcmin diameter part centered on \rxx. Data
  from MOS1 and MOS2 have been merged to produce the image. A few other faint
  and point-like sources, indicated by circles, are detected in the
  field of view.} \label{mos12_image}
 \end{figure}

\begin{figure}[t!!!]
 \psfig{file=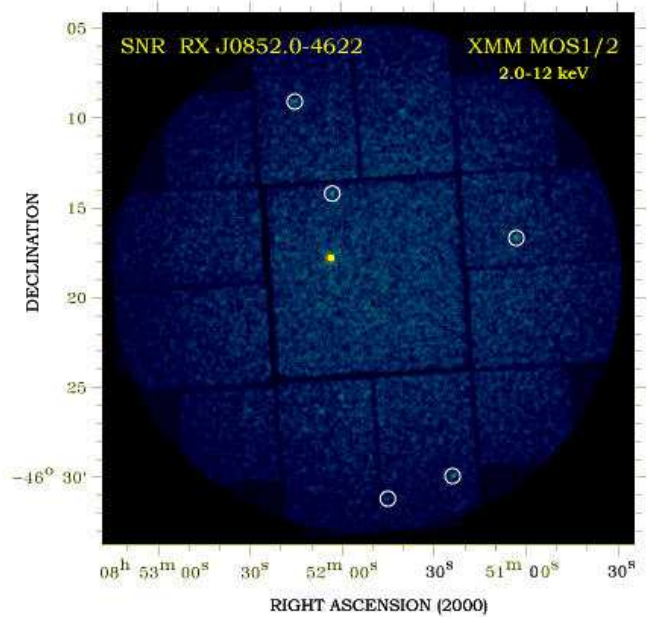,width=8.8cm,clip=}
 \caption[]{\renewcommand{\baselinestretch}{1.0}\small\normalsize\small
 The central part of \rxx\ in the energy band $2.0-12$ keV. Only very faint
  residual emission from the extended structure surrounding \cxx\ is visible.}
 \label{mos12_image_hard}
 \end{figure}

 The central part of \rxx\ was observed on 2001 April 27 by XMM-Newton in the course 
 of the guaranteed time program (Becker et al.~2002a). The exposure time was 
 $\sim 25$ ksec. In all observations, the MOS1/2 and the PN cameras were operated 
 in full-frame and extended full-frame mode, respectively. An image of the central 
 part of \rxx\ as seen with the MOS1/2 is displayed in Fig.\ref{mos12_image}. 
 \cxx\ is the brightest source in the field. It is located at the western edge 
 of a faint diffuse and irregularly shaped X-ray structure. The extended 
 emission was already visible in ASCA images, but less resolved from the neutron 
 star because of ASCA's wider point spread function. As the whole remnant emits 
 non-thermal X-ray emission along with a small thermal contribution from the 
 Vela remnant, it was not possible to conclusively identify the extended 
 structure in the ASCA images as plerionic (Slane et al.~2001). The XMM-Newton 
 observation thus allowed for the first time a somewhat more detailed 
 view of the field. The extended source has a size of $\sim 9' \times 14'$  
 and emits mostly in the soft band below $\sim 2$ keV. A MOS1/2 image showing 
 the field in the 2.0-12 keV band is displayed in Fig.\ref{mos12_image_hard}.
 No significant emission from the extended structure is visible beyond
 $\sim 2$ keV.

 Fitting model spectra to the data from \cxx, we tested the usual blackbody 
 and power-law models as well as models consisting of two blackbody components 
 and a blackbody and power-law component, respectively. With a reduced $\chi^2$ 
 of 0.952 (for 218 dof) the single blackbody model provides an excellent
 fit to the data, implying a column density of $N_H = (3.7 - 4.3)\times 10^{21} 
 \mbox{cm}^{-2}$, a temperature in the range $(4.28 - 4.50) \times 10^6$ K  
 and a formal emitting area of $R_{bb}=(0.31-0.36)$ km for an assumed 
 distance of 1 kpc. The unabsorbed X-ray flux and luminosity are $f_x=
 (1.82 - 2.96)\times 10^{-12}\,\mbox{erg cm}^{-2} \mbox{s}^{-1}$ and 
 $L_x(\mbox{\small d=1 kpc})=(2.17-3.53) \times 10^{32}\, \mbox{erg s}^{-1}$ 
 in the $0.5-10$ keV band, respectively. All parameters are given for a 90\% 
 confidence range.
 Compared with the blackbody fit, the power-law model is invalidated by a 
 larger reduced $\chi^2$ of 1.21 and residuals which indicate systematic 
 deviations from the measured data points. The fitted photon index of 
 $\alpha= 4.38 - 4.62$  is much steeper than what is usually found in 
 other young neutron stars which are well known for their non-thermal 
 emission. The blackbody fit, however, indicates a small unmodeled emission in 
 the hard band beyond $\sim 3$ keV. We therefore tested a compound 
 model consisting of a blackbody and an additional power-law component and
 fitted the non-thermal spectrum with a photon index of $\alpha = 2.85 \pm 1.5$.
 The flux and luminosity of these component are $f_x=3.7\times 10^{-13}
 \mbox{erg s}^{-1}\mbox{cm}^{-2}$ and $L_x(\mbox{d=1 kpc})= 4.4 \times 10^{31}$ erg/s.

 The temperature along with the small size of the emitting region
 invalidates the interpretation that the thermal radiation is emission from cooling of 
 the entire neutron star surface. If the blackbody model is the right description
 it points towards emission from a hot polar cap region on the neutron star surface.
 In the literature various models have been  proposed to produce hot spots on the
 neutron star surface. Almost all magnetospheric emission models predict a
 bombardment of the polar cap regions by energetic particles accelerated in
 the magnetosphere backwards to the neutron star surface (cf.~Cheng, Ho \& Ruderman
 1982; Harding \& Muslimov 2002 and references therein). Though \cxx\ is
 not known to be a radio pulsar it still could be possible that the hot polar
 caps are heated by particle bombardment while the radio emission itself is
 undetected due to an unfavorable beaming geometry. This is not an unlikely
 scenario. The opening angle of a radio beam is inversely proportional to
 the square root of the pulsar's rotation period so that the radio beams
 of slowly rotating pulsars can easily miss the observers line of sight and
 thus keep undetected.

 The heat transport in neutron stars is accomplished by electrons and
 positrons. A strong magnetic field is therefore expected to have an essential
 impact on the neutron star cooling as it channels the heat transport
 along the magnetic field lines and suppresses it in the perpendicular direction. 
 Neutron star cooling with a full treatment of a strong magnetic field
 thus should lead to an anisotropic heat flow and subsequently to an
 anisotropic surface temperature distribution, with the polar cap
 regions hotter than the
 equatorial regions.

 \begin{figure}[t!!!!]
 \psfig{file=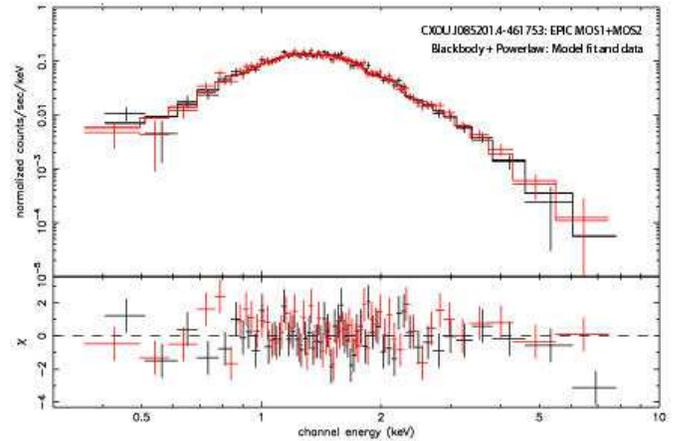,width=8.8cm,clip=}
 \caption[]{\renewcommand{\baselinestretch}{1.0}\small\normalsize\small
  The spectrum of \cxx\ obtained with XMM MOS1/2 and fitted with a blackbody 
  plus power-law model. The residuals of the fit are shown.} 
  \label{vela_jr_MOS12_spectrum}
 \end{figure}

 A great advantage of the CCD based EPIC detectors aboard XMM is that they
 not only provide spatial and high resolution spectral information but 
 have observing modes which are fast enough for timing studies even of 
 millisecond pulsars (Becker et al.~2002; Kuster et al.~2002, Kendziorra et 
 al.~2002). During the observation of \cxx\ the MOS1/2 and PN cameras were 
 operated in full frame and extended full frame mode, providing a temporal 
 resolution of 2.6 s and 200 ms, respectively. While this is not fast enough 
 to search for coherent pulsations on a millisecond time scale, it allows to 
 search the signal for periodicities in a period range in which anomalous X-ray  
 pulsars (AXPs) are seen.  Application of  fft-based search algorithms along 
 with a more detailed periodicity search around possible frequency candidates, 
 however, did not result in a detection of a significant periodic signal. 

\begin{figure}[t!!!!]
  \centerline{\psfig{file=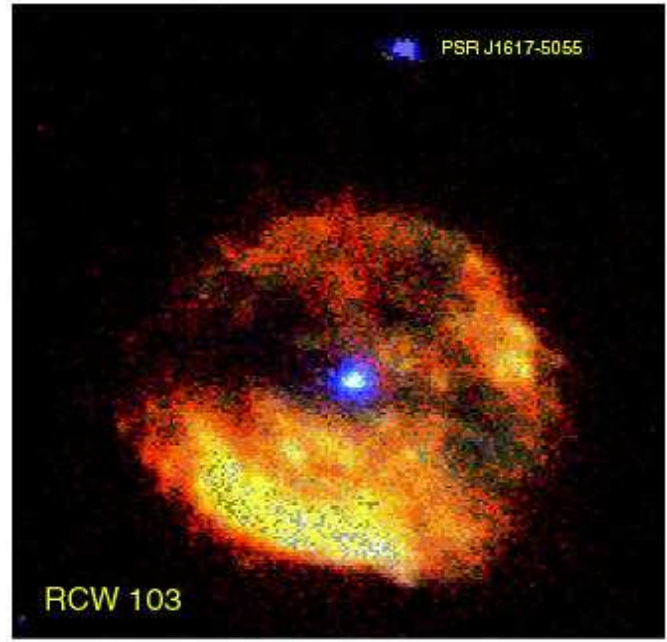,width=8.8cm,height=8.6cm,clip=}}
  \caption[]{\renewcommand{\baselinestretch}{1.0}\small\normalsize\small
   XMM MOS1/2 image of the supernova remnant RCW 103. The red color represents photons
   in the $0.3-0.75$ keV band, green and blue correspond to $\sim 0.75-2$
   keV and $2-10$ keV, respectively. The neutron star candidate is the hardest X-ray source
   in the remnant. The 69 ms pulsar PSR J1617$-$5055 is located 7 arcmin off the supernova
   center. The remnant extent is $\sim 10'$. \label{xmm_rcw_103}}
  \end{figure}

 Another example of a radio silent neutron star in a supernova remnant is
 1E~161348$-$5055, the projected location of which is in RCW 103 (cf.~Fig.\ref{xmm_rcw_103}). 
 This source was discovered with the Einstein Observatory (Tuohy \& Garmire 
 1980) and has an estimated age of $\sim 2000$ yrs. The remnant is mostly  
 known for its central point source, which was considered for many years 
 to be the prototype of a cooling neutron star. No X-ray pulsations and no  
 radio or optical counter part could be detected. ROSAT observations 
 suggested that the source emits hard X-rays (Becker et al.~1993). 
 Comparing two ASCA observations of RCW 103 Gotthelf, Petre 
 \& Vasisht (1999) found an order-of-magnitude decrease in luminosity 
 over four years, which suggests that this object may be an accreting source 
 rather than a cooling neutron star. Likewise puzzling is the $\sim 6$h 
 periodicity of its flux reported by Garmire et al.~(2000) from Chandra
 observations and archival ASCA data. 

 \begin{figure}[h!!!!!]
  \centerline{\psfig{file=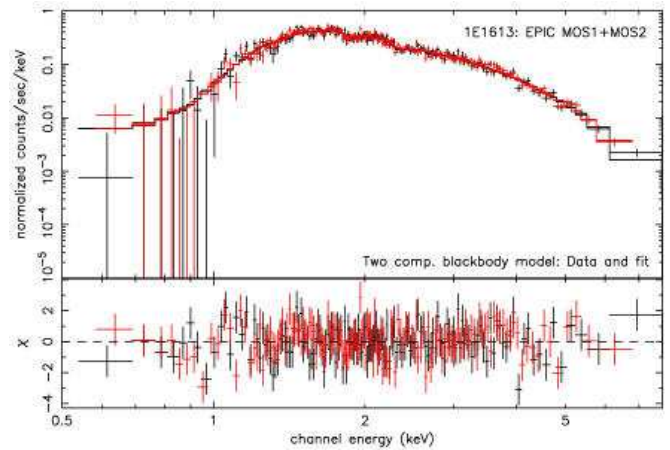,width=8.8cm,clip=}}
  \caption[]{\renewcommand{\baselinestretch}{1.0}\small\normalsize\small
   X-ray spectrum of 1E~161348$-$5055 in RCW 103 obtained from XMM MOS1/2 data.
   The spectrum is fitted by a two component blackbody model. Red and black data
   points are from the MOS1 and MOS2 cameras, respectively. A small spectral feature
   is seen at $\sim 0.98$ keV.\label{xmm_rcw_103_spec}}
  \end{figure}

 \begin{figure}[t]
  \centerline{\psfig{file=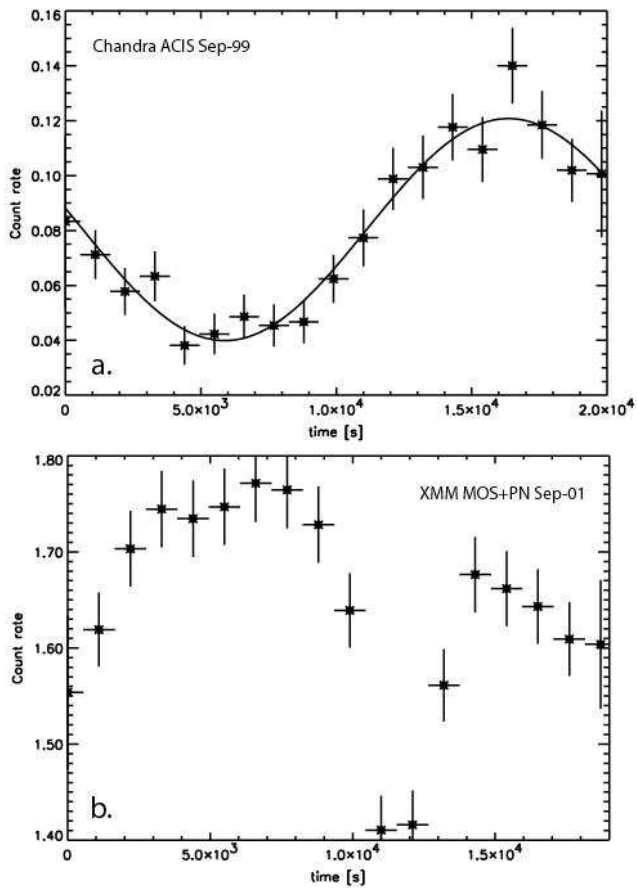,width=8.8cm,clip=}}
  \caption[]{\renewcommand{\baselinestretch}{1.0}\small\normalsize\small
   X-ray lightcurve of 1E~161348$-$5055 observed with Chandra (a.) and XMM-Newton (b.)
   in September 1999 and 2001, respectively. The Chandra lightcurve indicates a sinusoidal
   modulation with a period of $\sim 6$h. The lightcurve observed by XMM-Newton shows for
   the first time clear indication of an eclipse and thus identifies 1E~161348$-$5055 as
   a neutron star in a binary system (Becker 2001).} \label{xmm_rcw_103_lightcurve}
  \end{figure}

  XMM has observed RCW 103 in September 2001 for an exposure time of 20 ksec. 
  Figure \ref{xmm_rcw_103} shows a false color image of the remnant, derived from 
  the XMM MOS1/2 data.
  These data provide for the first time detailed spectral information on the 
  central point source. Chandra has observed the remnant in September 1999 
  and February 2001. Albeit superior imaging, small counting statistics in the
  first observation and severe pile-up effects in the second observation preclude a 
  meaningful spectral analysis.
  The energy spectrum of 1E~161348$-$5055 extracted from XMM MOS1/2
  data is shown in Fig.~\ref{xmm_rcw_103_spec}. The status 
  of the detector response matrix which is available for the EPIC-PN small-window
  mode and which is still preliminary, did not allow to include this data in 
  the analysis up to now, but 
  will provide additional spectral information in the  $0.3-10$ keV band.
  Fitting various spectral models to the MOS data a simple blackbody was 
  found to describe the data almost up to $\sim 5$ keV, but leaving 
  unmodeled hard emission beyond. 
  A single power-law fit is acceptable on statistical grounds but requires 
  a photon spectral index  of $4.1 - 4.2$. Testing two component models, a double 
  blackbody is found to describe the data best ($\chi^2_\nu= 1.0$ for 287 dof), 
  resulting in $N_H= (1.5 - 1.8) \times 10^{22}\,\mbox{cm}^{-2}$, temperatures of 
  $T_1= (4.1 - 5.1) \times 10^6$ K and $T_2=(8.3 - 10) \times 10^6$ K 
  with $R^{bb}_1=1.7-3$ km and $R^{bb}_2=220-500$ m for the blackbody 
  radii of the emitting areas (assuming a distance of $d=3.3$ kpc). The 
  given numbers represent the 90\% confidence range. This result suggests 
  that the first thermal component is emitted from a rather large fraction
  of the surface, whereas the second component requires a 
  much smaller emitting area like a polar cap region. Compared with the surface 
  temperatures predicted by neutron star cooling models, $T_1$ is much 
  higher than what is  expected for a $\sim 2000$ yr young neutron star.

  In addition to the long term flux change the Chandra data suggested a 
  $\sim 6$h periodicity. Archival ASCA data were found to be consistent with
  this. However, a second Chandra observation taken in Feb 2000 did not really 
  confirm this periodicity and was more consistent with a $\sim 4$h period, 
  although 1E~161348$-$5055 was about ten times brighter 
  at that time than in September 1999. It was therefore of special interest 
  to see the lightcurve of
  1E~161348$-$5055 taken with XMM-Newton in September 2001. Figure 
  \ref{xmm_rcw_103_lightcurve} shows the lightcurve taken with Chandra and 
  XMM-Newton in September 1999 and 2001, respectively. The lightcurve taken
  by XMM-Newton does not show the $\sim 6$h periodicity but discovers an eclipse
  about 3h after the start of the observation, identifying 1E~161348$-$5055 as 
  the second accreting binary in a supernova remnant. The companion most likely 
  is a low-mass companion. Chandra observations taken in March 2002 confirm
  the XMM findings (Sanwal et al., 2002). The discovery of a candidate optical 
  counterpart in the near-IR was reported by Pavlov et al.~(these proceedings).
  
  Similar to the previous examples is the point source RX~J0822$-$4300
  in Puppis-A (cf.~Fig.\ref{Vela_Jr_Rosat}). With a distance of about 2.2 kpc 
  the remnant is further away than the Vela supernova remnant. The  
  kinematic age is estimated to $\sim 3500 - 4000$ years, and is therefore 
  a factor of about three younger than the Vela SNR. Discovered in Einstein data, 
  it became evident with 
  ROSAT that RX~J0822$-$4300 is most likely the neutron star expected to form  
  in the SN event (Petre, Becker \& Winkler 1996). As for \cxx\
  there is no radio emission detected from this source down to 0.75 mJy 
  and no optical counterpart could be detected 
  down to an observation limit of about 24 mag.

\begin{figure}[t]
  \centerline{\psfig{file=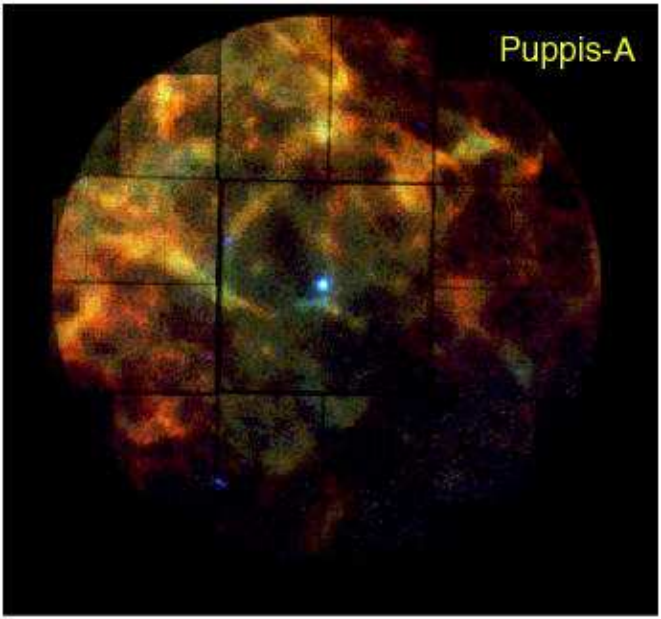,width=8.8cm,clip=}}
  \caption[]{\renewcommand{\baselinestretch}{1.0}\small\normalsize\small
   False color image of the central part of Puppis-A, depicting an area of 15 arcmin in
   radius centered on  RX~J0822$-$4300. The red color represents photons
   which are detected in the $0.3-0.75$ keV band, green and blue correspond
   to $\sim 0.75-2$ keV and $2-10$ keV, respectively. Note the amazing honeycomb
   structure seen in the remnant emission. \label{xmm_puppis}}
  \end{figure}

  In order to study the nature of RX~J0822-4300 in greater detail,
  the central part of Puppis-A was observed by XMM-Newton in April and 
  December 2001 for 20 kec and 30 ksec, respectively. Figure \ref{xmm_puppis}
  shows the false color image of the central part of the remnant taken with
  the two MOS1/2 cameras in April 2001. Like in RCW 103, the neutron star
  candidate is the hardest X-ray source in the remnant.

  The analysis of the  RX~J0822$-$4300 emission reveals an X-ray spectrum 
  similar to that observed for 1E~161348$-$5055. Fitting the data with a 
  simple blackbody model doesn't describe the spectrum beyond $\sim 3$ keV.
  A two component blackbody model and a model consisting of a blackbody
  and a power-law are fitting the spectrum equally well. The double blackbody 
  spectrum yields  $N_H= (3.6 - 4.4) \times 10^{21}\,\mbox{cm}^{-2}$, 
  temperatures of $T_1= (3.4 - 4)\times 10^6$ K and $T_2=(0.6 - 1.3) \times 
  10^7$ K with $R^{bb}_1=1.4-1.8$ km and $R^{bb}_2=40-310$ m for the blackbody 
  radii of the corresponding emitting areas (assuming a distance of $d=2.2$ kpc).
  A model consisting of a blackbody and a power-law yields similar results
  for the dominating blackbody component ($N_H= (2.9 - 3.6) \times
  10^{21}\mbox{cm}^{-2}$,  $T = (4.3 - 4.5) \times 10^6$ K, $R^{bb}=1.1 - 1.3$ km)
  but fits the hard emission with a power-law of photon-index $\alpha= 2.0 - 2.7$  
  (90\% confidence).  This slope is in agreement with what is observed for other 
  young pulsars showing magnetospheric emission.

  Zavlin et al.~(1999) have modeled the ROSAT data of  RX~J0822$-$4300 including 
  a hydrogen atmosphere on top of the neutron star. These models reduce the temperature of the
  thermal component by a factor of $\sim 2$ and require a somewhat larger
  size for the emitting region. Spectral fits which take into account models of a non-magnetic
  neutron star atmosphere of pure H or He are currently in progress. The X-ray luminosities 
  based on the two component models are in the range $L_x= (3-4) \times 10^{33}\,\mbox{erg/s}$.

  The EPIC-PN camera observed  RX~J0822$-$4300 in small-window mode 
  restricting the field of view to $4' \times 4'$ but with a time 
  resolution of 6 ms. Using archival ROSAT data of RX~J0822$-$4300  
  Zavlin et al.~(1999) found evidence for X-ray pulsations with a 
  period of $\sim 75.3$ ms. The XMM-Newton data, however, which are 
  of much better statistical quality than the ROSAT data do not 
  confirm this periodicity, leaving the rotation period of the  
  neutron star candidate unknown.

  From the observations of radio-silent compact objects in supernova
  remnants one can conclude that the emission properties of such 
  sources are quite different compared to ordinary radio pulsars.
  On the other hand, it is very plausible that they are 
  more common than radio pulsars, and the relatively small number 
  of the members of this class discovered so far is mainly due to 
  observational selection effects. It is much easier to detect 
  and identify active radio or X-ray pulsars than these ``radio-quiet'' 
  sources observable only in the X-ray band and located in a patchy
  X-ray bright supernova remnant environment.

 \begin{figure}[t!!!!!]
  \centerline{\psfig{file=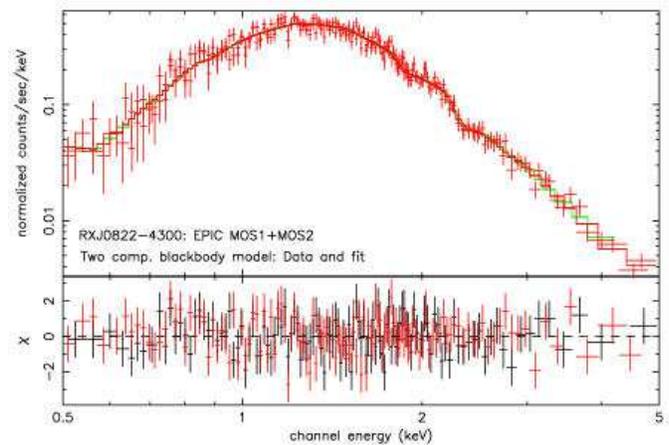,width=8.8cm,clip=}}
  \caption[]{\renewcommand{\baselinestretch}{1.0}\small\normalsize\small
   X-ray spectrum of the point source  RX~J0822$-$4300 in Puppis-A
   derived from XMM MOS1/2 data. The spectrum is fitted by a two component
   blackbody model. Red and black data points represent the data taken with
   the MOS1 and MOS2 cameras, respectively. The residuals of the fits shown below
   shown amount to a reduced $\chi^2_\nu$ value of 0.9444 for $\nu=244$
   degrees of freedom. \label{xmm_puppis_spec}}
  \end{figure}

\section{The cooling neutron stars} \label{cooling_ns}

 As discussed in \S\ref{photospheric}, soft X-ray radiation of rotation-powered pulsars
 of an age between $\sim 10^5-10^6$ yrs should be dominated by emission 
 from the cooling hot neutron star surface. These pulsars are old enough for their magnetospheric 
 emission to become fainter than the thermal surface emission, but they are still young 
 and hot enough to be detectable in the soft X-ray range. There are three middle-aged 
 pulsars, i.e. Geminga, PSR B0656+14 and B1055$-$52, from which thermal X-ray radiation from 
 the surface of the cooling neutron star has certainly been observed. Because of the 
 similarity of their emission properties, they were dubbed  as {\em the three Musketeers}
 by Becker \& Tr\"umper (1997).  
 It follows from the ROSAT and ASCA observations of the brightest middle-aged pulsar
 B0656+14 that the thermal component cannot be described by a single temperature
 blackbody spectrum, i.e.~the neutron star surface temperature is not uniform 
 (Greiveldinger et al.~1996). In a simple model the thermal component consists  
 of a soft thermal component (TS) from most of the neutron star surface 
 (at $E\lapr 0.5-1$ keV) and a hard thermal component (TH) from hot polar caps,
 heated by relativistic particles, for instance. Alternatively, the temperature 
 non-uniformity can be due to an anisotropic heat conductivity in the neutron 
 star crust caused by the anisotropic magnetic field --- the heat flux across 
 the field is reduced  whereas it is preferred in directions perpendicular to the 
 field, so that the magnetic poles are expected to be hotter than the equatorial 
 regions (Potekhin \& Yakovlev 2001).
 The thermal component dominates in the UV through the soft X-ray range (up to 
 $1-2$ keV), whereas the non-thermal component described by a power-law 
 (PL) spectrum prevails in the IR, optical, hard X-ray and gamma-ray bands.

 PSR 1055-52 and Geminga are not as luminous as PSR 0656+14. Their spectra
 measured by ROSAT and/or ASCA are well described by a two component model
 consisting of a blackbody (TS) and a non-thermal (PL) component. Whether
 a second thermal spectral component (TH) was not
 detected because of poor counting statistics or whether the component is intrinsically
 absent has been a pending question which prompted observations with XMM-Newton.

\begin{figure}[t]
  \centerline{\psfig{file=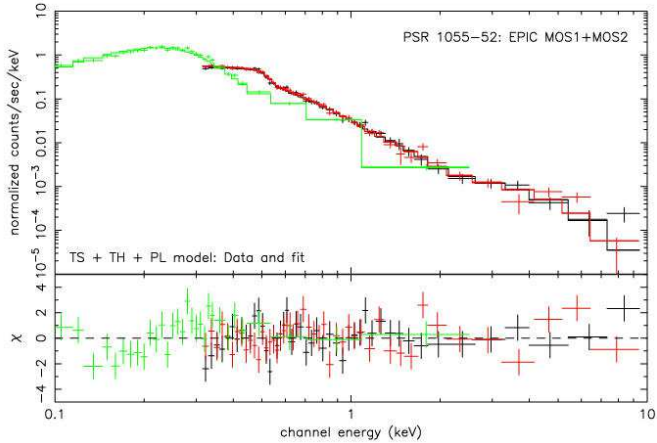,width=8.8cm,clip=}}
  \caption[]{\renewcommand{\baselinestretch}{1.0}\small\normalsize\small
   X-ray spectrum of PSR 1055$-$52 derived from ROSAT and XMM MOS1/2 data. As for PSR
   0656+14 the spectrum requires three components consisting of two thermal
   (TS+TH) and a harder non-thermal power-law component (PL). The green points
   represent ROSAT data. Red and black points represent MOS1 and MOS2
   data, respectively. \label{xmm_1055_spec}}
  \end{figure}

 PSR 1055-52 was observed by XMM-Newton for about 96 ksec in December 2000. The  
 X-ray spectrum is shown in Figure \ref{xmm_1055_spec}. It is interesting to 
 note that the two component spectral model which describes the ROSAT and ASCA 
 data does not fit the pulsar spectrum observed by XMM-Newton (Becker 2001). 
 Including the ROSAT data in the spectral analysis, 3 spectral components are 
 required to fit the spectrum in the wide band from 0.1$-$10 keV. The pulsar's 
 spectral properties are thus indeed found to match those observed in PSR 0656+14.
 The soft thermal component (TS) yields a blackbody temperature of $T_{TS}=
 (7.1 \pm 0.3)\times 10^5$ K, similar to what was obtained from the ROSAT data 
 with a somewhat smaller error, though. The second and harder thermal component (TH)
 is described by  $T_{TH}=(1.4 \pm 0.1)\times 10^6$ K. The corresponding blackbody 
 emitting areas are $R_{TS}=31 \pm 2 $ km and $R_{TH}=2.6 \pm 0.3$ km, 
 respectively, assuming a distance of 1.5 kpc. The non-thermal component 
 (PL) is described by a power-law with a photon index of $\alpha=1.89\pm 0.32$.
 The column absorption is $N_H=(2.1 - 2.4)\times 10^{20}\,\mbox{cm}^{-2}$
 (all errors represent 90\% confidence limits). For the unabsorbed flux in the 
 $0.5-10$ keV band we obtain for each of the three components 
 $f_{x}^{TS} = 2.3\times 10^{-13}\, \mbox{erg s}^{-1}\mbox{cm}^{-2}$, 
 $f_{x}^{TH} = 2.1 \times 10^{-13}\,\mbox{erg s}^{-1}\mbox{cm}^{-2}$ and 
 $f_{x}^{PL} =  1.4 \times 10^{-13}\,\mbox{erg s}^{-1}\mbox{cm}^{-2}$. The corresponding
 bolometric luminosities are $L_{TS}=1.74\times 10^{33}\,\mbox{erg s}^{-1}$
 and  $L_{TH}=1.9\times 10^{32}\, \mbox{erg s}^{-1}$. For the luminosity of
 the non-thermal component we obtain $L_{x}= 3.8 \times 10^{31}\, \mbox{erg s}^{-1}$.
 More detailed results on the pulsar spectrum as well as the results of a phase 
 resolved spectral analysis are given in Becker et al.~(2002b).

 It should be stressed that the inferred effective temperatures, and hence the
 radius-to-distance ratios, depend on the model used to describe the thermal
 component. For instance, if one assumes that the neutron star surface is covered by a
 hydrogen or helium atmosphere, the effective temperatures are lower than
 those derived from the simple blackbody fits by a factor of $1.5-3$
 (Pavlov et al.~1995). Atmospheres dominated by heavy elements give temperatures
 close to the blackbody temperatures. However, the heavy element
 atmosphere spectra should show numerous absorption lines and photo-ionization
 edges (Rajagopal \& Romani 1996; Zavlin et al.~1996). Recent observations
 by Chandra and XMM-Newton of PSR 0656+14 (Marshall \& Schulz 2001) as well
 as of the neutron stars RXJ 1856.5-3754 (Burwitz et al.~2001) and RX J0720.4-3125 
 (Paerels et al.~2001) have not shown any sign for line emission from a neutron star 
 atmosphere but are perfectly consistent with a pure blackbody like emission.

  The X-ray pulses of PSR 1055$-$52 were discovered by \"Ogelman \& Finley (1993) using
  ROSAT data. The temporal emission properties were found to indicate a  phase shift of 
  about $100^\circ$ at $0.5-0.6$ keV accompanied by an apparent increase in the pulsed  
  fraction from $\sim 10\%$ to $\sim 65\%$. Although the narrow bandwidth of ROSAT
  along with a decreasing sensitivity beyond 1 keV resulted in large errors of the 
  energy dependent pulsed fraction and phase shift, it was suggested that this dichotomy 
  is associated with the thermal and non-thermal emission components seen in the 
  pulsar's X-ray spectrum. The non-thermal magnetospheric emission is expected to be  
  stronger beamed than the thermal one. Magnetospheric emission thus could show up with 
  a pulsed fraction of up to 100\%, depending on the viewing geometry. 

 \begin{figure}[t]
  \centerline{\psfig{file=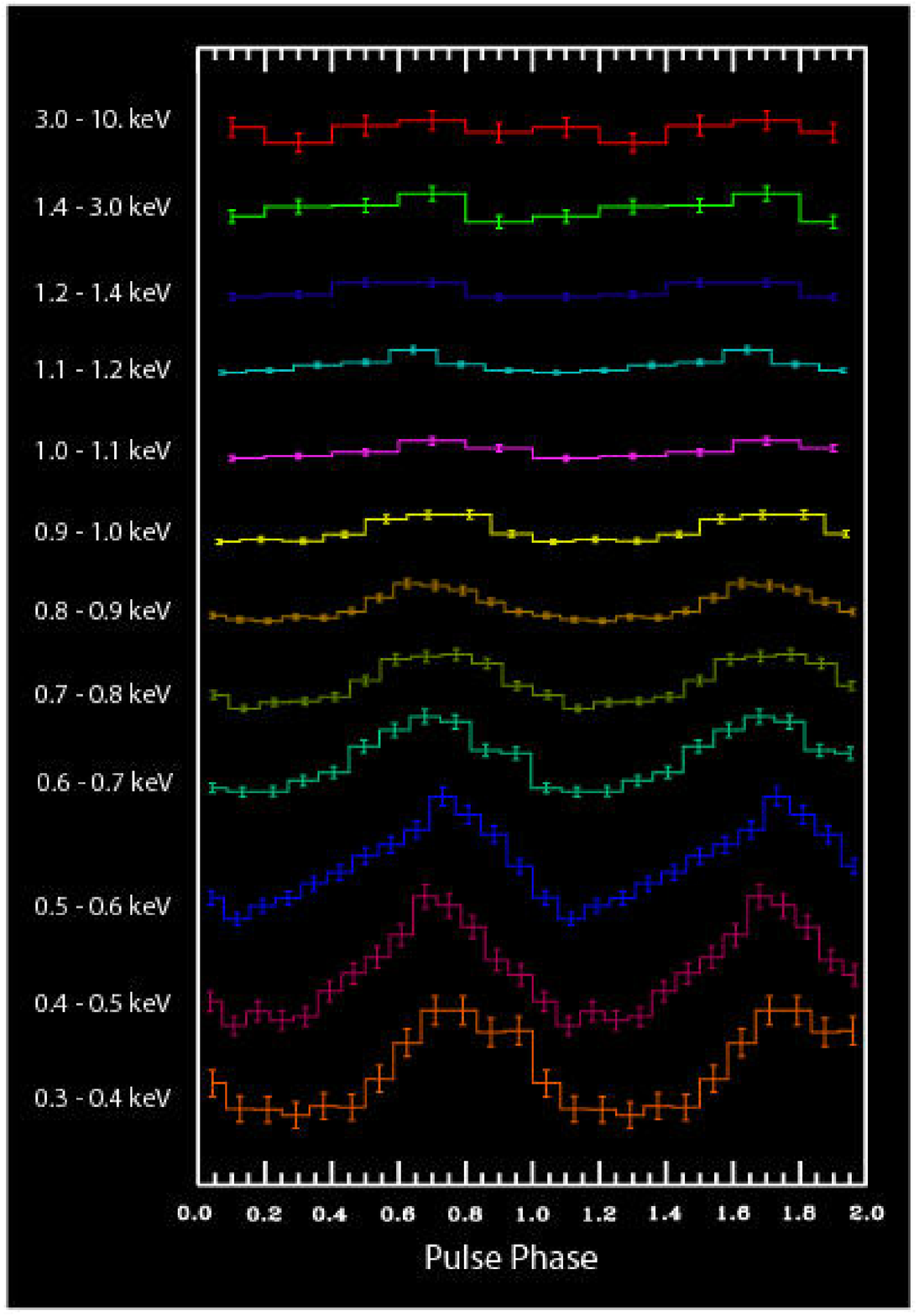,width=8.8cm,clip=}}
  \caption[]{\renewcommand{\baselinestretch}{1.0}\small\normalsize\small
   X-ray pulse profiles of PSR 1055$-$52 for different energy bands. The pulse profile
   is sinusoidal. Two complete cycles are shown for clarity. The energy dependent phase shift
   in the $0.3-3$ keV band is not as significant as suggested from ROSAT observations.
   \label{xmm_1055_lc}}
  \end{figure}

  XMM-Newton has observed PSR 1055-52 with the EPIC-PN camera set up in timing mode. 
  As described in the  introduction (cf.~\S \ref{introduction}), this mode is very 
  sensitive to any enhanced background emission which is accumulated over the whole CCD 
  and condensed along with the source signal into a 1D-image. After barycentering the 
  photon arrival times and correcting for the satellite orbit, the timing signal is 
  detected with high significance.  The X-ray pulse profiles, which are shown for 
  different energy bands in Fig.\ref{xmm_1055_lc}, are all approximately sinusoidal. 
  A reason for this sinusoidal shape could be the non-uniformity of the 
  surface temperature due to the presence of a strong magnetic field 
  of  $\sim 5\times 10^{12}$ G, as discussed above.

 \begin{figure}[t]
  \centerline{\psfig{file=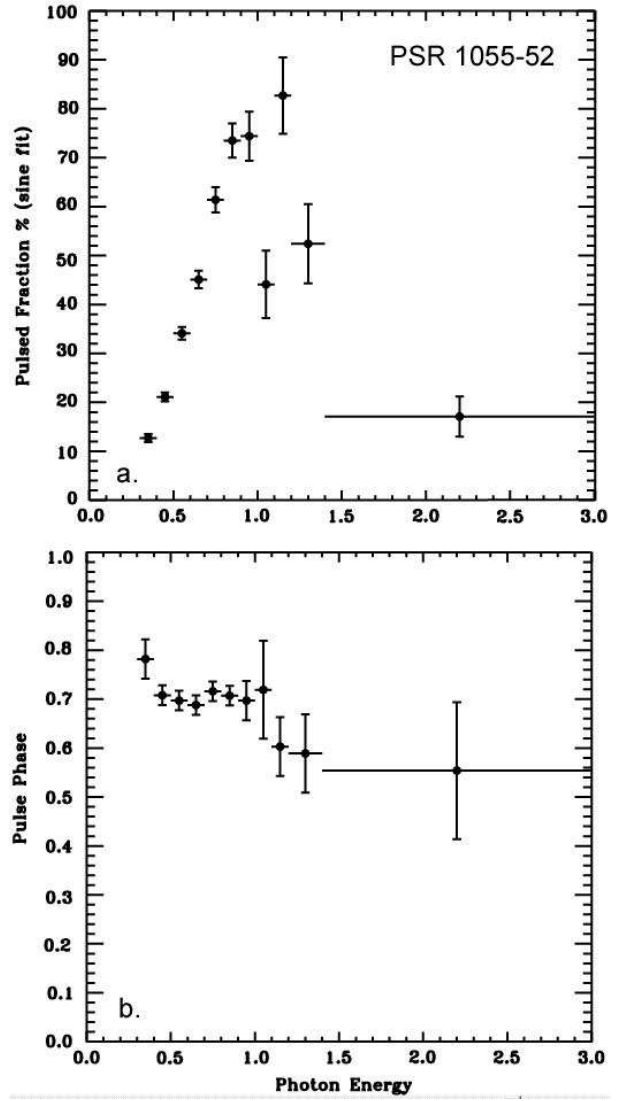,width=8cm,clip=}}
  \caption[]{\renewcommand{\baselinestretch}{1.0}\small\normalsize\small
  {\bf{a.}} Fraction of pulsed flux vs.~photon energy observed for PSR 1055$-$52
   by XMM-Newton. The pulsed fraction peaks at $\sim 1.1$ keV and decreases beyond that
   energy. The decrease goes along with a decreasing signal-to-noise ratio which we suggest to 
   be caused by the fading pulsar signal at higher energies in combination with a high
   DC level from soft-proton flares which are present over the whole
   observation. Note the dip in the pulsed-fraction at $\sim 1$ keV. {\bf{b.}}
   Pulse phase vs.~photon energy. The XMM-Newton data are in agreement with no
   significant phase shift in the 0.4-3 keV band. A significant pulsed  signal is
   seen up to $\sim 3$ keV only. \label{xmm_1055_pf_und_phase_vs_E2}}
  \end{figure}

  Figure \ref{xmm_1055_pf_und_phase_vs_E2}a shows the pulsed  fraction $P_f$ as a function
  of photon energy. $P_f$ increases from $\sim 15\%$ in the $0.3-0.4$ keV band up to
  $\sim 80\%$  in the $1.1-1.2$ keV band.  The fraction of the pulsed flux at higher
  energies is undetermined in the EPIC-PN data. The observed decrease at
  $E \ge 1.3$ keV is due to an enhanced background signal, caused by soft proton
  flares which contaminate the observation and which can not be corrected for. The
  high background together with a fast decline of the pulsar emission beyond $\sim 2$
  keV causes a strong reduction of the signal-to-noise ratio in the hard energy band,
  preventing the detection of X-ray pulsations in the EPIC-PN data from most of 
  the non-thermal (PL) emission component. 
  The phase angles of the X-ray pulses observed in different energy bands is shown 
  in Figure \ref{xmm_1055_pf_und_phase_vs_E2}b. The XMM data indicate that the energy 
  dependence of the pulse phase is by far not as strong as indicated by ROSAT. This
  can also be seen in the pulse profiles shown in Fig.~\ref{xmm_1055_lc}.

  Since all active pulsars are powerful sources of strong winds, one should 
  expect that they generate pulsar-wind nebulae (PWNe), similar to those observed 
  around the Crab-like and Vela-like pulsars. The PWN sizes should scale as 
  $(\dot{E}/p_0)^{1/2}$, where $p_0$ is the pressure of the ambient medium. The 
  existence of X-ray bright PWNe (albeit of much larger sizes) around several 
  pulsars, including the three musketeers, was reported by Kawai \& Tamura (1996) 
  based on ASCA observations. In addition, the suggestion of a clumpy 
  pulsar-wind nebula near to PSR B1055$-$52 was made by Shibata et al.~(1997). 
  The authors proposed  the "clumps" of X-ray emission to be created by the interaction 
  of a pulsar-wind outflow with the  local interstellar environment. The very low 
  signal-to-noise ratio of the "clumps", however, limited any detailed analysis 
  based on just the ASCA data. In an analysis combining BeppoSAX and ROSAT data, Becker 
  et al.~(1999) confirmed the existence of X-ray sources surrounding PSR 1055$-$52 
  but came to a different conclusion with respect to its interpretation. With ROSAT,
  and partly with BeppoSAX, the clumpy X-ray emission was resolved as the superposition
  of a number of point sources, well separated from the pulsar and unlikely to be associated 
  with it. Several of these sources were found to have optical and radio counterparts
  using the DSS2 and a radio image taken by Stappers et al.~(1999) at 1.4 GHz. 

 \begin{figure}[t]
  \centerline{\psfig{file=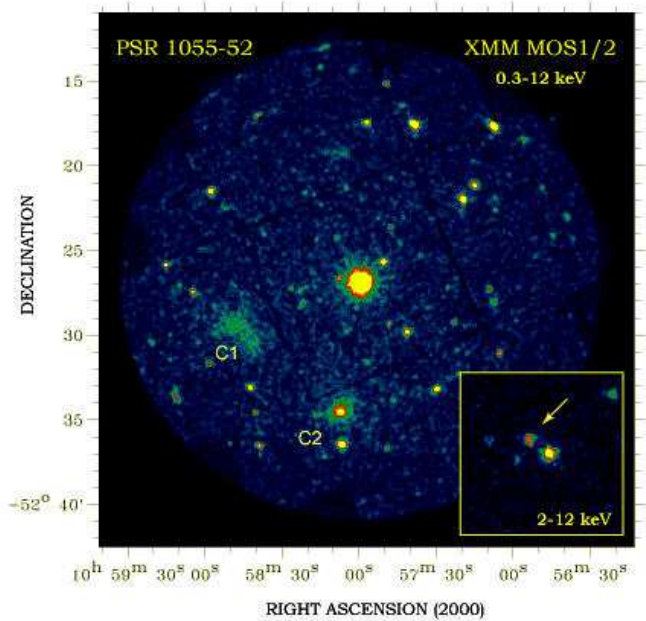,width=8.8cm,clip=}}
  \caption[]{\renewcommand{\baselinestretch}{1.0}\small\normalsize\small
   The X-ray sky around PSR 1055-52 as seen by the MOS1/2 detectors in the $0.3-12$ keV band.
   The inset in the lower-right corner is a zoomed view on the central $4'\times 4'$ field
   around PSR 1055-52 for 2-12 keV. The arrow indicates the X-ray source XMMU J105801.8-522639
   which is located $\sim 30"$ offset from  PSR 1055-52 which is the brightest source in the image.
   \label{xmm_1055}}
  \end{figure}

  The ROSAT data in the $1.0-2.4$ keV band were quite limited as far as the 
  photon statistics of the X-ray sources neighboring PSR 1055$-$52 is concerned. 
  It was therefore of special interest so see this sky region observed by XMM-Newton,  
  which provides a sensitivity about a factor of 10 higher than ROSAT. The X-ray   
  sky around PSR 1055-52 as seen by  XMM-Newton in the $0.3-12$ keV band is 
  shown in Fig.\ref{xmm_1055}. Apart from the many X-ray sources detected in 
  the field the source denoted by Becker et al.~(1999) as C1 appears to be  
  extended by about $1\times 4$ arc-minutes. Its location 
  coincides with that of four faint radio point sources some of which have 
  optical counterparts detected in the DSS2 (Becker et al.~1999). Similar
  results are found for the other putative X-ray clump C2. 
  Interesting to note is an X-ray source located  30" offset from the pulsar 
  position (cf.~the inset in Fig.\ref{xmm_1055}). This source, denoted as 
  XMMU J105801.8-522639, shows up only in the hard band and was not recognized 
  in any other observation of PSR 1055-52 so far. Inspecting the data from the 
  XMM-Newton optical monitor we found that XMMU J105801.8-522639 has no optical counter 
  part down to $m_v\sim 22$. The radio image of Stappers et 
  al.~(1999) shows a faint radio source near to PSR 1055-52 which roughly 
  matches with the position of XMMU J105801.8-522639. It might be the 
  radio counterpart of XMMU J105801.8-522639.

\section{The millisecond Pulsars}  \label{ms_psrs}

  By the middle of 2002, about 100 millisecond pulsars have become known, out of 
  which 57 are located in the galactic plane (Camilo 1999; Lommen et al.~2000;
  Manchester et al.~2000). The others are in globular clusters (Kulkarni
  \& Anderson 1996; Camilo et al.~2000, D'Amico et al.~2001) which provide
  a favorable environment for the recycling scenario (Rasio, Pfahl \& Rappaport
  2000). Only 10 of the 57 galactic plane ms-pulsars are solitary
  (including PSR B1257+12 which has a planetary system); the rest are in
  binaries, usually with a low-mass white dwarf companion. 

  Millisecond pulsars were studied exclusively in the radio domain until
  the early 1990's when PSR B1957+20 was detected by ROSAT (Kulkarni et
  al.~1992; Fruchter et al.~1992). Unfortunately only a handful of counts were
  recorded from this object so that no detailed data analysis was possible.
  The identification was based on the positional coincidence with the radio
  pulsar. The first ms-pulsar with X-ray pulsations was PSR J0437$-$4715
  (Becker \& Tr\"umper 1993). Since then further detections followed
  (Halpern 1996; Verbunt et al.~1996; Danner et al.~1997; Saito et al.~1997; 
  Takahasi et al.~1999; Becker et al.~1998a; Becker et al.~1998b; Becker 
  et al.~1999; Mineo et al.~2000, Becker et al.~2000, Grindlay et al~2002)
  which three years after the launch of Chandra and XMM-Newton sum up to 
  almost 50\% of all X-ray detected rotation-powered pulsars. Table 
  \ref{ms_psr_sym_tab} provides an up-to-date summary.

  Although for most of these objects no detailed spectral and temporal information 
  is available, the number of detected counts allows at least an estimate of the X-ray 
  flux and luminosity assuming a standard 
  X-ray spectrum  (cf.~Tab.\ref{ms_psr_sym_tab}). XMM-Newton will observe
  several of these pulsars during AO1 (cf.~Tab.\ref{ao1_targets}) and hopefully will 
  provide the required spectral and temporal information to finally identify the
  emission mechanism responsible for the observed X-ray emission. Chandra and HST
  observations of the ms-pulsars detected in 47TUC are scheduled 
  for October 2002 so that also for these objects more detailed information will 
  become available in the near future.
  For the remaining pulsars, data of six objects provide sufficient spectral and/or temporal 
  information to identify the origin of their X-ray emission (Becker \& Tr\"umper 
  1999 and references therein; Sakuray et al.~2001; Mineo et al.~2000).  
  Observations by Chandra and XMM-Newton have provided the missing spectral 
  information for some of them only recently (Kuiper et al.~2002; Zavlin et 
  al.~2002, Becker et al.~2002c).

 \begin{table*}[ht!!!!!!!!!!!!!!!!!!!!!!!!!!!]
 \begin{picture}(130,230)(10,50)
 \put(0,0){\psfig{figure=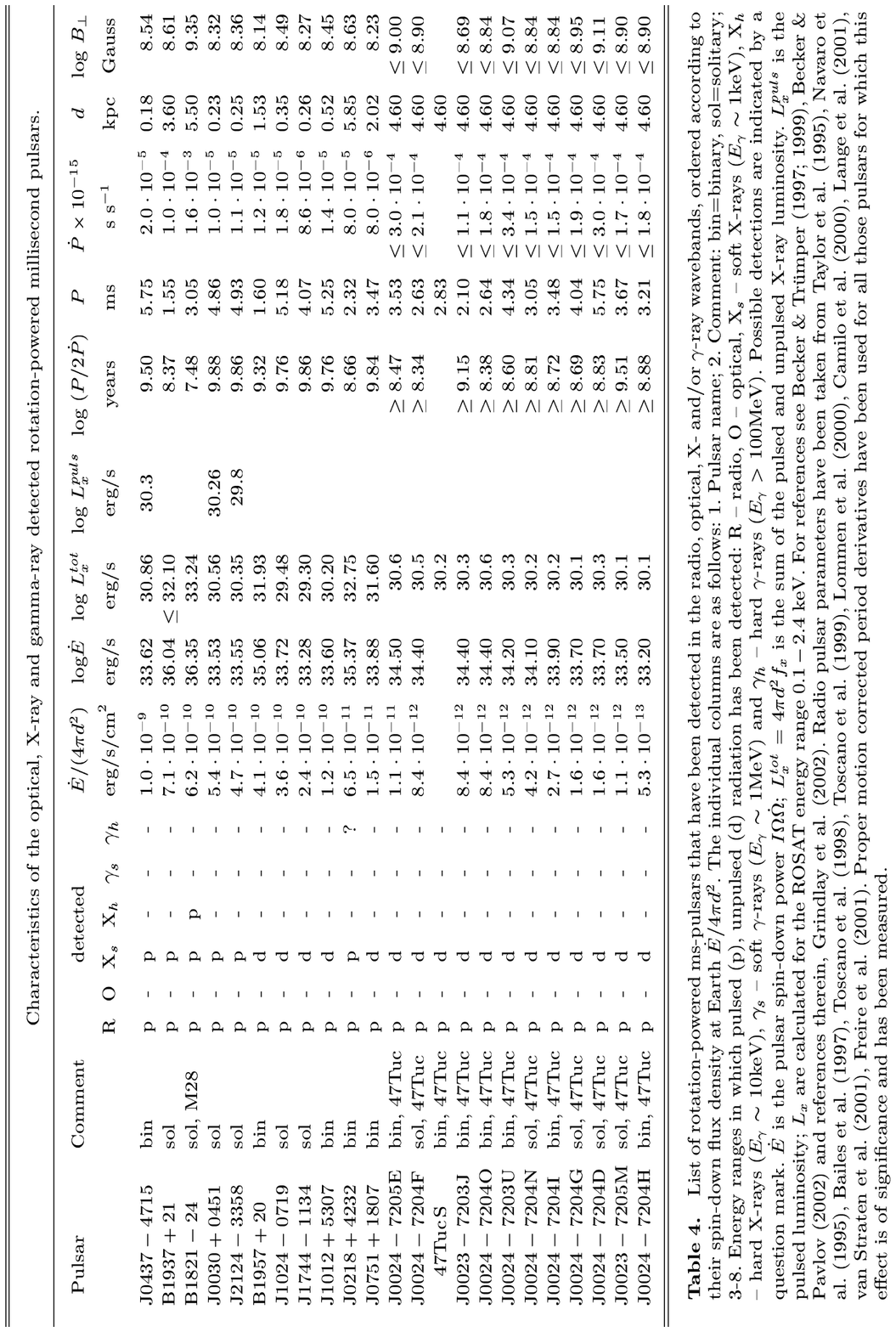,width=24cm,clip=}}
  \end{picture}
  \refstepcounter{table} \label{ms_psr_sym_tab}
 \end{table*}

\begin{figure}[t]
  \centerline{\psfig{figure=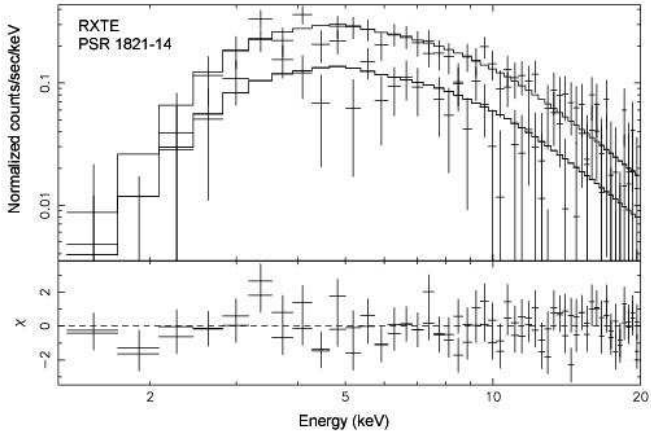,width=8.8cm,clip=}}
  \caption[]{\renewcommand{\baselinestretch}{1.0}\small\normalsize\small
  X-ray spectrum of the millisecond pulsar PSR 1821-24 extracted from XTE data. 
  A power-law spectrum yields a valid description of the data of up to 20 keV
  (Saito, priv.~com.).}\label{xte_1821}
  \end{figure}

\begin{figure*}[t!!!!!!!!!!!!!!!!!!!!!!!]
  \centerline{\psfig{figure=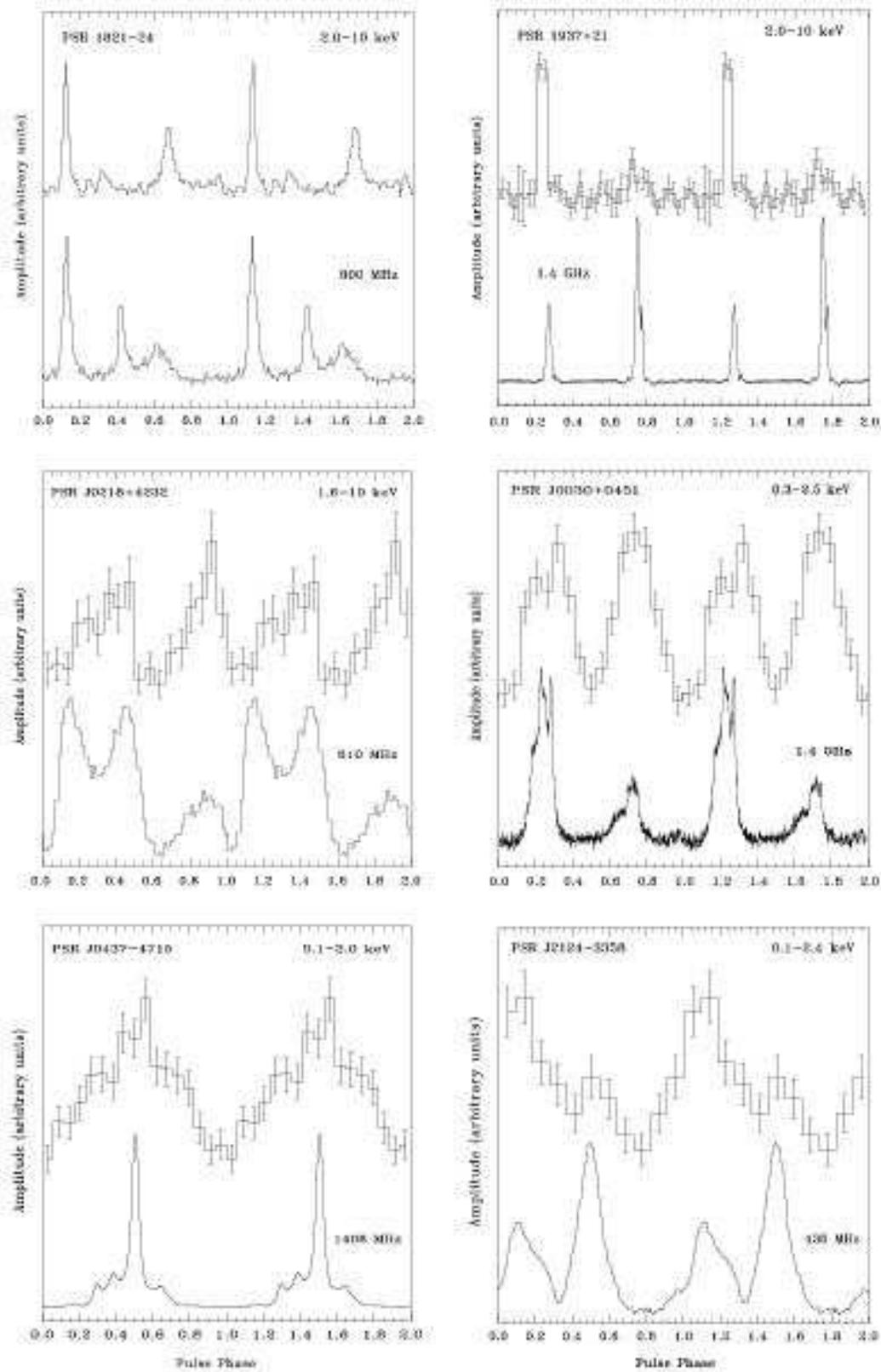,width=13.9cm,clip=}}
  \caption[]{\renewcommand{\baselinestretch}{1.0}\small\normalsize\small
   Integrated lightcurves for all ms-pulsars for which X-ray pulses have been detected.
   The upper panel phase histograms show the pulse profiles for the energy bands noted.
   The radio profiles are shown in the lower panel for comparison. Two full pulse cycles
   are shown for clarity. The relative phase between the radio and X-ray pulses is only known
   for PSR 1821-24, B1937+21, 0218+4232 and PSR J0437-4715. The  phase alignment in all
   other cases is arbitrary due to the lack of accurate clock calibration.}
   \label{ms_profiles}
   \end{figure*}

  The X-ray emission observed from PSR B1821$-$24, PSR B1937+21 and
  PSR J0218+4232 ($P\sim 1.5-3$ ms, $\log\dot{E}\sim 35-36$ erg s$^{-1}$)
  is produced by non-thermal processes (Kawai \& Saito 1999, Takahashi et
  al.~1998; 2001, Mineo et al.~2000, see also Nicastro et al.~2002, these  
  proceedings). This is concluded from their power-law spectra (photon-index
  $\alpha\sim 1.5-2$) and pulse profiles with narrow peaks and pulsed 
  fractions of up to $\sim 90\%$ (cf.~Fig.~\ref{ms_profiles}).  
  PSR B1821$-$24 was detected by RXTE up to $\sim 20$ keV (cf.~Fig.~\ref{xte_1821}). 
  All these pulsars show relatively hard X-ray emission, which made it 
  possible to study some of their emission properties already with ASCA, 
  BeppoSAX and RXTE.
  For the remaining pulsars ($P\ge 4$ ms, $\log\dot{E}\sim 33-34$ erg s$^{-1}$)
  the X-ray emission is found to be much softer, which in the absence of definite
  spectral information led to the conclusion that in these pulsars thermal polar-cap 
  emission dominates. The argument for this second group being purely thermal 
  emitters was that their pulse profiles are  broad, and that the pulsed fraction 
  is small ($\sim 20-30\%$) in comparison to the non-thermal  sources 
  (pulsed fraction $\ge 50-60\%$). The argument is not unreasonable --  thermal 
  polar-cap emission should result in broad sinusoidally modulated soft emission 
  with a pulsed fraction $\le 50\%$ (Pavlov \& Zavlin 1998), whereas magnetospheric 
  emission is  expected to create narrow pulse-peaks and large pulsed 
  fractions.

  But, of course, given the limited amount of data the physical interpretation 
  is not fully justified and therefore not at all unique. 
  Becker \& Tr\"umper (1997; 1999) have emphasized that the radiation 
  cone confining the magnetospheric emission is likely to yield sharp peaks at one aspect 
  angle but may be less well defined when viewed from other directions producing 
  a less strong modulation. Thus 
  a broad, weakly modulated pulse profile does not provide unambiguous evidence for 
  thermal emission. Indeed, based on the 
  $L_x$ vs $\dot{E}$ relation (most of the X-ray pulsars lie close to the line 
  $L_x \sim 10^{-3} \dot{E}$, cf.~Becker \& Tr\"umper 1997) 
  one can argue on phenomenological grounds that the X-ray emission 
  from all ms-pulsars is dominated by magnetospheric emission. The gross similarity 
  between the X-ray and radio pulse profiles is part of this argument. 

  For the nearest and brightest millisecond pulsar, PSR J0437$-$4715, it was
  already evident in the ROSAT and ASCA data that the X-ray emission consists 
  at least of two different spectral components. Either a two temperature thermal 
  model or a model composed of a blackbody ($T\sim 1.2 \times 10^6$ K) and a power-law
  (photon-index $\alpha=2.9$) fitted these data but also a broken power-law spectrum 
  gives a valid description (Becker \& Tr\"umper 1999). The emitting area deduced from 
  the blackbody fits was found to be very small ($R_{bb}\sim 510\,\mbox{m}\,(d/180\,
  \mbox{pc})$), implying the existence of relatively small hot-spots on the 
  neutron star surface from which the X-rays are emitted (Becker \& Tr\"umper 1997; 1999). 

  Chandra observed PSR J0437$-$4715 in May 2000 (Zavlin et al.~2002, cf.~also 
  Zavlin et al.~and Pavlov et al.,~these proceedings). These data, together with
  the ROSAT data, provide the most significant spectral information available 
  from this pulsar by now and suggest the presence of a  power-law component 
  and two thermal components in the pulsar's X-ray spectrum. The thermal 
  components are interpreted as emission from a hot polar cap, having a 
  non-uniform temperature distribution with a hot core ($T_{core} = 2.1 
  \times 10^6$ K, $R_{core}=0.12$ km) and a cooler rim ($T_{rim}=0.54 
  \times 10^6$ K, $R_{rim}\sim 2.0$ km). The power-law component yields a 
  photon index of $\alpha\sim 2.2$. The size of the polar cap is found to 
  be roughly in agreement with the theoretical predictions. Defined as the 
  area of open field lines in which the bombardment by relativistic particles 
  is expected, it is $R_{pc}=R(R\,\Omega/c)^{1/2}$. Assuming $R=10$ km for the 
  neutron star radius and $\Omega=1.09\times 10^3$ for the pulsars angular 
  frequency $R_{pc}=1.9$ km. It is interesting to mention that the pulsar
  spectrum measured by Chandra is still consistent with a broken power-law with 
  properties similar to those fitted to ROSAT and ASCA data by Becker \& Tr\"umper 
  (1999); $N_H= 0.3^{+0.5}_{-0.2}\,\mbox{cm}^{-2}$, $\gamma_1=2.0\pm 0.2$, 
  $\gamma_2=3.6\pm 0.1\,\mbox{keV}$ below and above the break energy 
  $E_b=1.1\pm 0.1$ keV. Zavlin et al.~(2002) argue against the broken power-law  
  model because of an apparent discrepancy between the flux estimated from 
  this model and observed by EUVE DSI - despite a known discrepancy in the 
  cross-calibration between ROSAT and EUVE. If the broken power-law model 
  is relevant, and this is not excluded presently by the Chandra data, it would 
  imply that the purely 
  non-thermal pulsar spectrum would bend down significantly in the X-ray band, 
  which could be due to a 
  deficit of high-energy charged particles in the pulsar's magnetosphere. 
  
\begin{figure}[t]
  \centerline{\psfig{file=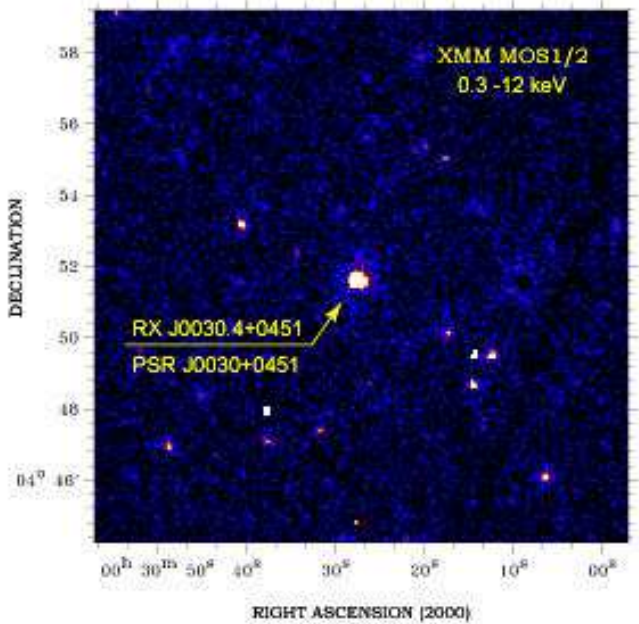,width=8.8cm,clip=}}
  \caption[]{\renewcommand{\baselinestretch}{1.0}\small\normalsize\small
   The X-ray sky around PSR J0030+0451 as seen by the MOS1/2 detectors in the
   $0.3-12$ keV band. The position of PSR J0030+0451 is marked. There is
   no evidence for any extended emission. \label{xmm_0030_image}}
  \end{figure}

  A pulsar with spin parameters similar to those observed in PSR 0437$-$4715 
  is the solitary ms-pulsar PSR 0030+0451 (cf.~Table \ref{ms_psr_sym_tab} and 
  Fig.\ref{xmm_0030_image}).  X-ray emission from this pulsar was discovered 
  during the final ROSAT PSPC observations (Becker et al.~2000). Though the 
  "dying" PSPC detector had lost most of its spectral capability 
  a timing analysis revealed strong X-ray pulsations 
  with a pulsed fraction of $69 \pm 18\%$. This and the gross similarity 
  between the double peaked X-ray pulse profile and the Crab-like
  radio profile taken at 1.4 GHz suggested that the X-rays from  PSR 0030+0451
  were most likely of non-thermal origin, although its spin-parameters appear
  very similar to those of PSR J0437$-$4715 suggesting a soft pulsar
  spectrum  (Becker et al.~2000). It was therefore important to observe this 
  pulsar with XMM-Newton in order to identify the emission mechanism 
  by the spectrum rather than from just the pulse profile
  and the similarity of the spin-parameters with PSR J0437$-$4715.

  PSR 0030+0451 was observed by XMM-Newton in 2001 June $19-20$ for a duration 
  of $\sim 30$ ksec. The MOS1/2 and the PN cameras were operated in full-frame 
  and fast-timing mode, respectively. The latter mode required to constrain the
  satellite roll angle to  $67-68$ degrees in order to prevent source confusion
  in the PN data (cf.~\S\ref{introduction}). The thin filters where used 
  for the MOS1/2 and the PN cameras in order to block stray light and optical 
  leakage from bright foreground stars. The observation is affected by 
  several soft-proton flares covering the whole observation. Cleaning the 
  data for the times of excessive background reduced the effective exposure 
  time to 17.8 ksec. For the spectral analysis we selected all events in 
  the MOS1/2 within a circle of 38 arcsec radius, containing $\sim 87\%$ 
  of the encircled energy of a point source. The X-ray flux and luminosity 
  were corrected accordingly. The data are not affected by pile-up.

  \begin{figure}[t]
  \centerline{\psfig{figure=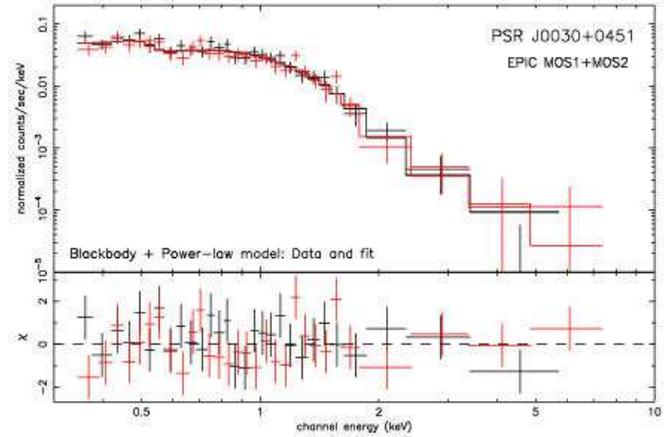,width=8.8cm,clip=}}
  \caption[]{\renewcommand{\baselinestretch}{1.0}\small\normalsize\small
  X-ray spectrum of the millisecond pulsar PSR J0030+0451 extracted from XMM data.
  The read and black points correspond to data extracted from the MOS1 and
  MOS2 detectors, respectively. The data are fitted by a composite model consisting
  of a blackbody and power-law model. The residuals of the fits are shown at the bottom.}\label{xmm_0030}
  \end{figure}

  Figure \ref{xmm_0030} shows the energy spectrum of PSR 0030+0451 observed 
  with XMM-MOS1/2. The spectrum is rather flat in the soft band. The pulsar 
  is detected up to $\sim 7$ keV. Motivated by the results obtained for  
  PSR J0437$-$4715 we tested various thermal and non-thermal spectral models 
  such as blackbody and power-law models and combinations of them  as well 
  as broken- and curved power-law spectra. 
  A single blackbody model did not yield a valid description of the spectrum because of 
  severe systematic deviations and a rather large reduced $\chi^2$ of $\chi^2_\nu
  =1.4$ (for 56 dof).
  In comparison to this, a single power-law model fits the observed spectrum
  ($\chi^2_\nu=0.92$ for 56 dof) but requires a high column absorption 
  of $N_H=(2.1 - 3.46)\times 10^{21}\,\mbox{cm}^{-2}$ with a rather steep power-law 
  slope of $\alpha=(4.2 - 5.1)$ for the photon-index (90\% confidence range). 
  For comparison, the column density along the pulsar's line of sight 
  through the entire Galaxy is just $N_H= 3\times 10^{20}\,
  \mbox{cm}^{-2}$ (Dickey and Lockman 1990).  Models consisting of a combination 
  of a blackbody and a power-law as well as of a two   
  component blackbody model all yield an acceptable description of the observed 
  spectrum. For the blackbody plus power-law model ($\chi^2_\nu=0.87$ for 54 dof) 
  we obtain  $N_H=(0-2.5)\times 10^{20}\,\mbox{cm}^{-2}$, $T = (2.2 - 2.7) \times 
  10^6$ K, $R_{bb}=(50 -100)$ m and a photon index of $\alpha=2.65-3.45$ 
  (90\% confidence range). The pulsar distance was assumed to be 0.23 kpc 
  (Lommen et al.~2000). The corresponding total flux and luminosity in the 
  0.1--2.4 keV band is $f_x \sim 5.7 \times 10^{-13}\,\mbox{erg s}^{-1}
  \mbox{cm}^{-2}$ and $L_x \sim 3.6 \times 10^{30}\,\mbox{erg s}^{-1}$. 
  For the 0.5--10 keV band we compute $f_x \sim 1.9 \times 10^{-13}\,
  \mbox{erg s}^{-1}\mbox{cm}^{-2}$ and $L_x \sim 1.2 \times 10^{30}\,
  \mbox{erg s}^{-1}$, respectively.
  The two component blackbody model fits the spectrum with comparable
  quality and yields $N_H=(0-1.9)\times 10^{20}\,\mbox{cm}^{-2}$, $T_{bb1}  = 
  (2.6 - 3.7) \times 10^6$ K, $T_{bb2}  = (1 - 1.8) \times 10^6$ K, 
  $R_{bb1}=(20 - 60)$ m and $R_{bb2}= (130 - 340)$ m. 
  For the X-ray conversion efficiency in the 0.1-2.4 keV band we find $L_x/\dot{E}
  =1.1\times 10^{-3}$, in agreement with the general trend found by 
  Becker \& Tr\"umper (1997). In addition to these composite models we tested 
  a broken power-law and a curved power-law spectrum (Fossati et al.~2000)  
  either of which provides an  excellent fit to the data ($\chi^2_\nu =0.93$ 
  for 55 dof and $\chi^2_\nu =0.92$ for 51 dof for the broken and curved power-law 
  model, respectively). Fixing the column absorption to zero, we find the spectral  
  break point at $E_b=(0.97 - 1.28)$ and photon indices of $\alpha_1 =  
  (1.98 - 2.4)$ and  $\alpha_2 =  (3.88 - 5.36)$, respectively. The 
  unabsorbed flux and luminosity obtained from this model for the 0.1--2.4 
  keV band are $f_x \sim (4.3 - 6.8) \times 10^{-13}\,\mbox{erg s}^{-1}
  \mbox{cm}^{-2}$ and $L_x \sim (2.7-4.3) \times 10^{30}\,\mbox{erg s}^{-1}$,
  respectively. For the 0.5-10 keV band we find  $f_x \sim (1.5 - 2.1) 
  \times 10^{-13}\,\mbox{erg s}^{-1} \mbox{cm}^{-2}$ and $L_x \sim 
  (0.95-1.3) \times 10^{30}\,\mbox{erg s}^{-1}$. The fitted parameters 
  are very similar to those found in the analysis of the Chandra data 
  of PSR J0437-4715. Broken power-law fits thus provide an attractive 
  alternative to describe the X-ray spectrum of PSR J0437-4715 and 
  PSR J0030+0451 in a pure non-thermal scenario.  More detailed results 
  on the spectral analysis are presented in Becker et al.~(2002c).

 \begin{figure}[t!!!!!!!!]
  \centerline{\psfig{figure=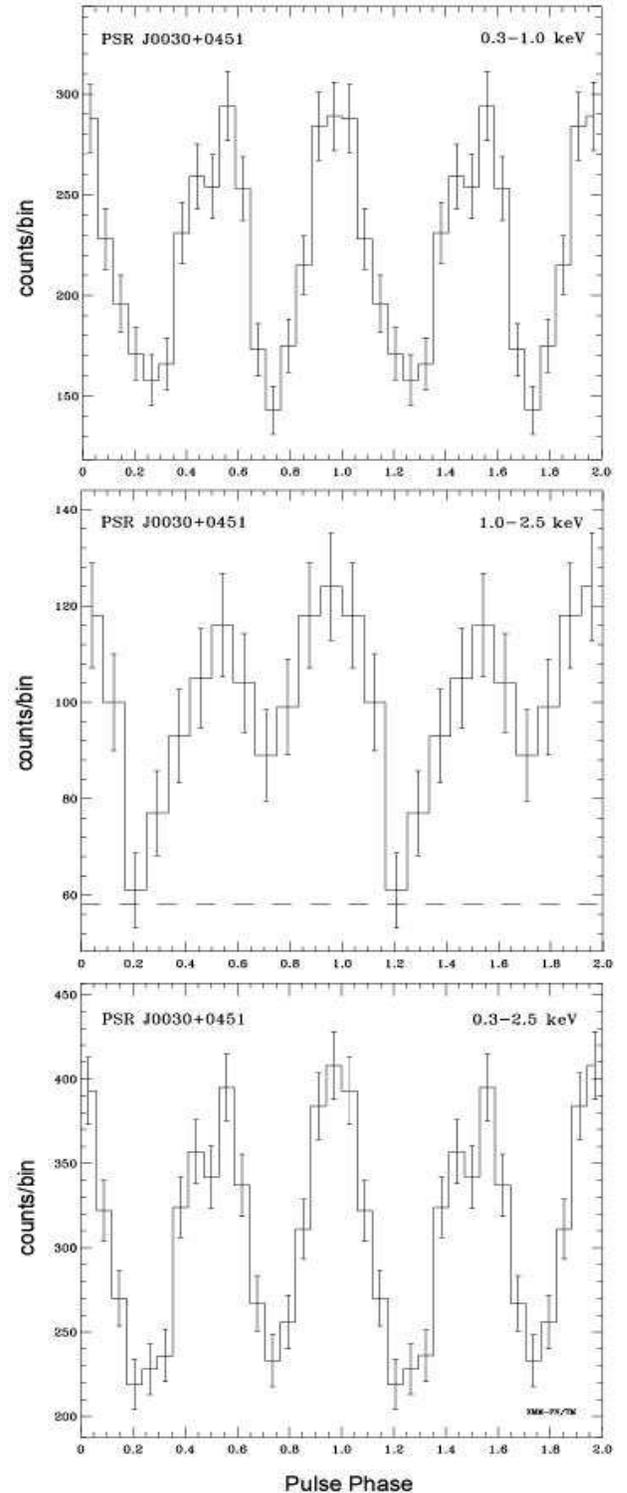,width=8.2cm,height=20cm,clip=}}
  \caption[]{\renewcommand{\baselinestretch}{1.0}\small\normalsize\small
   Pulse profiles of PSR J0030+0451 observed by XMM-Newton.  Two full pulse cycles
   are shown for clarity. The profile appears double peaked with a pulsed fraction
   of $50 \pm 3 \%$ in the $0.3-2.5$ keV band.\label{xmm_0030_lcs}}
  \end{figure}

  The rotation period of PSR J0030+0451 is 4.86 ms. The temporal resolution of the
  EPIC-PN camera in timing mode is 0.03 ms and thus is sufficient to measure the pulse
  profile and pulsed fraction as a function of photon energy. To extract the events from 
  PSR J0030+0451 we selected an area of $8 \times 199$ pixel from the central CCD \#4
  of the PN camera. According to the fractional encircled energy this means that we 
  sample about 70\% of the point-spread function in $x$- and 100\% in $y$-direction. 
  Applying the usual corrections for the barycentering and the satellite orbit, a 
  strong pulsed signal is detected in the $\sim 0.3-2.5$ keV band. Below and beyond
  that bandwidth the cumulative instrument and sky-background contribution per pixel
  exceeds that of the pulsar signal which prevents a detection at those energies (see 
  Becker et al.~2002c for further details). The pulse profiles obtained from events 
  in the 0.3--1.0 keV, 1.0--2.5 keV and 0.3--2.5 keV bands are displayed in 
  Fig.\ref{xmm_0030_lcs}. Each lightcurve shows two rather narrow peaks which are 
  separated in phase by $\sim 0.5$. XMM thus confirms our findings obtained in the 
  final ROSAT observation (Becker et al.~2000). The pulsed fraction is found to
  to be $(49 \pm 3)\%$ in the $0.3-1.0$ keV and $(66 \pm 12)\%$ in the $1.0-2.5$ keV 
  band (errors are $1\sigma$) using a bootstrap method (Swanepoel et al.,1996). 
  For the $0.3-2.5$ keV bandwidth we obtain $51\pm 3\%$. The gross 
  similarity between the X-ray and radio pulse profiles becomes even more striking with 
  these new pulse profiles obtained with XMM. As can be seen in Fig.\ref{ms_profiles},
  the peak separation of the radio and  X-ray lightcurves of  PSR J0030+0451
  appear to be in agreement with each other, suggesting a close relation between 
  the X-ray and radio emission mechanism. It is evident that this holds for all
  the other pulse profiles shown in the chart. It is straight forward to
  understand this gross similarity in terms of a common non-thermal emission mechanism
  but it is hard to understand how a mixture of thermal and non-thermal emission mechanisms
  can produce this similarity. In the case of PSR J0437-4715, for which the {\em broad} 
  X-ray pulse profile often is taken as an additional argument for the thermal origin
  of its soft X-radiation, it should be noted that also the radio emission has a duty cycle
  of $\sim 80\%$ so that the radio pulse itself appears rather broad. 

  In the light of these new results it will be interesting to see the data 
  from more millisecond pulsars observed by XMM-Newton and Chandra in order to 
  further explore these interesting objects and their emission mechanisms. 

\section{Concluding remarks}

  In spite of the first impressive achievements summarized above, lots of 
  questions still remain. First, the evolution of neutron stars, beginning with 
  their violent birth in supernova explosions, is far from being well 
  understood. Until recently, a common prejudice had been that all 
  neutron stars are born as active, fast spinning rotation-powered pulsars, 
  which slow down their rotation, eventually stop their activity and, after 
  crossing a ``death line'', get into the ``pulsar graveyard''. A former
  pulsar remains in the graveyard forever, cool and quiet, unless it is 
  captured by a flying-by star (e.g., in a globular cluster) and forms a 
  close binary, where accretion onto the neutron star can spin it up 
  (recycle) to periods that short that it again becomes an active pulsar.
  However, as the results from XMM-Newton and Chandra have shown, it is likely
  that the picture is not as simple as that.  In particular, it appears that several 
  very young neutron stars (like those in Puppis-A and RCW 103) are not active 
  pulsars at all. Others are found to have spin periods longer than the canonical 
  limit of 4 seconds. Were those objects born with such a slow spin-period or 
  did they spin down rapidly after their formation because of the presence of 
  an ultra-strong magnetic field? Or is it simply an unfavorable beaming which 
  precludes to see the pulsed emission?
  Since such objects are not seen in radio, and are extremely faint in optical,
  they could not be observed until the onset of the X-ray astronomy era, which
  means that our perception of neutron star early evolution was strongly
  biased in favor of much easier observable radio pulsars. Why are
  many (perhaps, the majority of) nascent neutron stars not active pulsars? Is
  it because they are indeed magnetars, whose super-strong magnetic field
  inhibits the pulsar activity? Or, on the contrary, are their magnetic fields
  so weak and/or is their rotation so slow that the pulsar does not turn on?
  Or is their pulsar activity quenched by accretion of debris of the supernova
  explosion? Are the (apparently young) anomalous X-ray pulsars and soft
  gamma-ray repeaters indeed magnetars or are their unusual observational
  properties due to quite different reasons, like a residual disk?

  Observing middle aged neutron stars at X-ray energies has provided valuable
  information on the pulsars' surface temperature for some objects. The XMM
  data have provided further evidence that the surface temperature
  distribution is not uniform and several, at least two,  thermal components are required to
  describe the observed spectra. The question remains what mechanisms are
  responsible for the non-uniform surface temperature distribution? Is the
  presence of the strong magnetic field already sufficient to channel the
  heat flow in the star to produce the observed  non-uniformity,
  or is it due to the presence of polar caps which are heated by a bombardment
  of energetic particles accelerated in the neutron star's magnetosphere?
  Other questions are related to the chemical composition of the neutron star
  surface layer and the presence of a neutron star atmosphere. Why does none of
  the data taken so far show any direct evidence for the existence of a
  neutron star atmosphere except 1E1207-5209 (cf.~Pavlov et al., these
  proceedings; Mereghetti et al., 2002 and references therein)? No emission 
  or absorption features are observed in the LETG spectrum taken from the 
  nearby isolated neutron star RX J1856-3754, but a blackbody was measured   
  with very high accuracy (Burwitz et al.~2001; cf.~Pavlov et al.,these 
  proceedings).

  Crucial for building a consistent evolutionary scenario of pulsar emission properties
  at X-ray energies are older pulsars in the age bracket $10^6-10^7$ years.  
  What is the dominant emission process producing their X-rays? Only a few of 
  these objects turn out to be bright enough to be observable at all. 
  None of them has been observed by XMM-Newton or Chandra so far but a few 
  objects are 
  scheduled for XMM's AO1 (cf.~Tab.\ref{ao1_targets}).  These pulsars are 
  of particular importance for the study of particle acceleration and high 
  energy radiation processes on the neutron star surface and in the 
  magnetosphere. This is because of their outstanding location  in the 
  $P-\dot{P}$ parameter space, where they are found in the area of strong magnetic 
  fields ($\sim 10^{11}-10^{12}$ G) but spin-down ages which are intermediate 
  between the middle aged cooling neutron stars and the very old 
  $\sim 10^9-10^{10}$ years millisecond pulsars. Will the thermal
  emission observed in cooling neutron stars simply fade away with increasing
  age due to the neutron star cooling or will the star be kept hot at about
  $(0.5-1) \times 10^5$ K over millions of years due to energy dissipation by
  processes such as internal frictional heating ($\dot{E}_{diss}\sim10^{28}
  -10^{30}\,\mbox{erg/s}$) and crust cracking or by vortex pinning and creeping?

  As far as millisecond pulsars are concerned, what are the pulsar parameters
  which determine the relative strength of the thermal and non-thermal emission
  components? Or is the pulsar emission non-thermal for all detected objects? 
  Why do the pulse profiles observed in the radio and X-ray 
  bands show this strong gross similarity, even when their X-ray
  emission appears to be emitted mainly from heated polar-caps like proposed 
  for PSR J0437-4715? What is causing the relation between the emission
  properties and energy spectra, which is  observed in the radio and X-ray band?

  In this paper we have shown that observatories like XMM-Newton provide
  very exciting new results on all classes of neutron stars and that this
  observatory will be able to provide the answer for many of the open questions
  mentioned above. We are confident that XMM-Newton and Chandra will have a 
  similar strong impact on our understanding of neutron star emission
  mechanisms as ROSAT had few years ago. With respect to this it might be worth
  to mention that we still use the soft photons detected by ROSAT to explore
  the X-ray spectra of pulsars because none of the new instruments can compete in
  the soft band down to energies of 0.1 keV.

\begin{acknowledgements}
 We would like to thank the XMM team at ESTEC and MPE for their help with the data 
 analysis, especially Dirk Grupe for the compilation of the data analysis cookbook 
 and Michael Freyberg for his help during the data processing and analysis. 
 Uwe Lammers, Eckhard Kenziora, Markus Kuster, Tim Oesterbroek and Ed Serpell 
 for the help with the clock calibration. WEB is thankful for discussion with 
 M.Weisskopf and J.Grindlay.
\end{acknowledgements}

\clearpage


\begin{thebibliography}{}

\bibitem[]{} Aschenbach, B., 1998, Nat, 396, 141

\bibitem[]{} Aschenbach, B., Iyudin, A., Sch\"onfelder, V., 1999, A\&A, 350, 997

\bibitem[]{}  Bailes, M., Johnston, S., Bell, J.F., et al, 1997,  ApJ, 481, 386

\bibitem[]{} Becker, W., \& Tr\"umper, J. 1993, Nat, 365, 528

\bibitem[]{} Becker, W., Tr\"umper, J., Hasinger, G., Aschenbach, B, 1993, in {\em Isolated
 Pulsars}, eds. K.A. Van Riper, R.I. Epstein \& C. Ho, p116, Cambridge University Press

\bibitem[]{} Becker, W., Tr\"umper, J., 1997, A\&A, 326, 682

\bibitem[]{} Becker, W., Tr\"umper, J., 1998a, IAU Circular No. 6829

\bibitem[]{}  Becker, W., Tr\"umper, J., Hasinger, G., 1998b, IAU Circular No. 6845

\bibitem[]{}  Becker, W., Tr\"umper, J.  1999, A\&A, 341, 803

\bibitem[]{} Becker, W., Kawai, N., Brinkmann, W., Mignani, R., 1999, A\&A, 352, 532

\bibitem[]{} Becker, W., Tr\"umper, J., Lommen, A.N., Backer, D.C., 2000, ApJ, 545, 1015

\bibitem[]{} Becker, W., 2001, Talk presented at the second XMM Symposium  {\em New 
Vision of the X-ray Universe in the XMM-Newton and Chandra Era}', Nov. 2001, ESTEC

\bibitem[]{} Becker, W., Pavlov, G.G., 2002, in {\em The century of Space Science}, eds. J.Bleeker, 
J.Geiss \& M.Huber, Kluwer Academic Press, in press, astro-ph/0208356 

\bibitem[]{} Becker, W., Aschenbach, B., Iyudin, A., 2002a, A\&A, submitted

\bibitem[]{} Becker, W., et al., 2002b, A\&A, submitted

\bibitem[]{} Becker, W., et al., 2002c,  ApJ, submitted

\bibitem[]{}  Burwitz, V., Zavlin, V.E., Neuh\"auser, R., Predehl, P., Tr\"umper, J., Brinkman, A. C., 2001, A\&A, 379, 35L

\bibitem[]{} Camilo, F. 1999, in {\em Pulsar Timing, General Relativity and the Internal
 Structure of Neutron Stars}, ed. Z. Arzoumanian, F.~Van der Hooft, \& E.P.J.~van den Heuvel, 
Amsterdam: Koninklijke Nederlandse Akademie van Wetenschappen, p. 115

\bibitem[]{} Camilo, F., Lorimer, D.~R., Freire, P., Lyne, A.~G., \& Manchester, R.~N., 2000, ApJ, 535, 975

\bibitem[]{} Caraveo, P.A., Mignani, R., Pavlov, G.G., Bignami, G.F., 2002, in
{\em A decade of HST science}, Space Telescope Science Institute Symposium, eds.
M.Livio, K.Noll and M.Stiavell, p13

\bibitem[]{} Cheng, K.S., Ho, C. \& Ruderman, M.A., 1986, ApJ, 300, 500

\bibitem[]{}  Danner R., Kulkarni S.R., Saito Y., Kawai N., 1997, Nat 388, 751

\bibitem[]{}  D'Amico, N., Lyne, A.G., Manchester, R.N., Possenti, A., Camilo, F., 2001, ApJ, 548, L171

\bibitem[]{} Dickel, J.R., Carter, L.M., Memorie della Societa Astronomia Italiana, Vol. 69, p.845

\bibitem[]{}  Dickey J. M., Lockman F. J., 1990, ARA\&A., 28,215

\bibitem[]{}  Freire, P.C., Camilo, F., Lorimer, D.R., Lyne, A.G., Manchester, R.N., D'Amica, N.,
 2001, MNRAS, 326, 901

\bibitem[]{} Fruchter A.S., Bookbinder J., Garcia M.R., Bailyn C.D., 1992, Nat, 359, 303

\bibitem[]{} Fossati, G., Celotti, A., Chiaberge, M., et al., 2000, ApJ, 541, 166

\bibitem[]{} Garmire, G.P., Pavlov, G.G., Garmire, A.B., Zavlin V.E. 2000, IAU Circ. \#7350

\bibitem[]{} Green, A.,2000, {\em A Catalogue of Galactic Supernova Remnants}, Mullard Radio Astronomy Observatory, Cambridge, UK

\bibitem[]{} Greiveldinger, C., Camerini, U., Fry, W., et al. 1996, ApJ, 465, L35

\bibitem[]{} Grindlay, J.E., Camilo, F., Heinke, C.O., Edmonds, P.E., Cohn, H., Lugger, P., 2002, to apper in ApJ

\bibitem[]{} Gotthelf, E.V., Petre, R., Vasisht, G. 1999a, ApJ. 514, L107

\bibitem[]{} Halpern, J.P.,  Holt, S.S., 1992, Nat, 357, 222

\bibitem[]{} Halpern J.P., 1996, ApJ, 459, L9

\bibitem[]{} Harding, A.K., Muslimov, A.G., 2002, ApJ, 568, 862

\bibitem[]{} in't Zand, J.J.M., et al.,1998, A\&A, 331, L25

\bibitem[]{} Iyudin, A.F., Sch\"onfelder, V., Bennett, K., Bloemen, H., Diehl, R., Hermsen, W.,
Lichti, G.G., van der Meulen, R.D., Ryan, J., Winkler, C., 1998, Nat, 396, 142

\bibitem[]{}  Johnston S., Lorimer D.R., Harrison P.A., et al. 1993, Nat, 361, 613

\bibitem[]{}  Kawai, N.,  Tamura, K., 1996, in {\em IAU Colloquium 160}, eds S.Johnston,
 M.A.Walker and M.Bailes, p367

\bibitem[]{}Kawai, N., Saito, Y., 1999, Astro. Lett. and Communications, 38, 1

\bibitem[]{} Kaspi, V.M., Crawford, F., Manchester, R.N., Lyne, A.G., Camilo, F., D'Amico, N., Gaensler, B.M., 1998,
ApJ, 503, L161

\bibitem[]{} Kendziorra E., Kuster, M., Kirsch, M., Risse, P., Straubert, R., Becker, W., Str\"uder L.,
Treis, J., Lechner, P., Holl, P., 2002, SPIE Conference 4851,

\bibitem[]{} Kuiper, L., Hermsen, W., Verbunt, F., Ord, S., Stairs, I., Lyne, A., 2002, ApJ, in press, astro-ph/0206081

\bibitem[]{} Kulkarni, S.~R.,  Anderson, S.~B. 1996, in {\em Dynamical Evolution of
 Star Clusters -- Confrontation of Theory and Observations}: IAU
 Symposium 174, Kluwer Academic Publisher, p.181

\bibitem[]{}  Kulkarni S.R., Phinney E.S., Evans C.R., Hasinger G., 1992, Nat, 359, 300

\bibitem[]{} Kuster, M., Kendziorra, E., Benlloch, S., Becker, W., Lammers, U., Vacanti, G., Serpell, E.,
2002, in the Proceedings of {\em New Vision of the X-ray Universe in the XMM-Newton and Chandra Era}', Nov. 2001,
ESTEC, in press

\bibitem[]{}  Lange, C., Camilo, F., Wex, N., Kramer, M., Backer, D.C., Lyne, A.G., Doroshenko, O.,
 MNRAS, 326, 274

\bibitem[]{} Lyne, A.G., Jordan, C.A., Roberts, M.E., 2002, 
Jodrell Bank Crab Pulsar Timing Results, Monthly Ephemeris, University of Manchester

\bibitem[]{}  Lommen, A.N., Zepka, A., Backer, D.C., Cordes, J.M., Arzoumanian, Z.,
 McLaughlin, M., \& Xilouris, K. 2000, ApJ, 545, 1007

\bibitem[]{} Manchester, R.~N., Lyne, A.~G., Camilo, F., Kaspi, V.~M., Stairs, I.~H., Crawford, 
F., Morris, D.~J., Bell, J.~F., \& D'Amico, N., 2000, in {\it Pulsar Astronomy - 2000 and Beyond}, 
ed. M.Kramer, N.Wex, and R.Wielebinski, San Francisco: ASP, p.49

\bibitem[]{} Manchester at al., 2002, Catalog of pulsars, unpublished

\bibitem[]{}  Marshall, H.L., Schulz, N.S., 2001, AAS,  199, \#119.05 

\bibitem[]{} Mereghetti, S., 2001, ApJ, 548, L213

\bibitem[]{} Mereghetti, S., De Luca, A., Caraveo, P.A., Becker, W., Mignani, R., Bignami G.F., 2002, accepted for ApJ, astro-ph/0207296
 
\bibitem[]{}  Mineo, T., Cusumano, G., Kuiper, L., Hermsen, W., Massaro E.,
 Becker, W., Nicastro, L., Sacco, B., Verbunt, F., Lyne, A.G., Stairs, I.H., Shibata, S. 2000, A\&A, 355,  1053

\bibitem[]{}  Navaro, J., de Bruyn, G., Frail, D., Kulkarni, S.R., Lyne, A.G., 1995, ApJ, 455, L55

\bibitem[]{} \"Ogelman, H., Finley, J.P., 1993, ApJ, 413, L31

\bibitem[]{} \"Ogelman, H.B., 1995, in {\em The Lives of Neutron Stars},
eds M.Alpar et al., NATO ASI Series Vol. 450, Kluwer Academic Pblisher,   p101-120

\bibitem[]{} Paerels, F., Mori, K., Motch, C., Haberl, F., Zavlin, V.E., Zane, S., Ramsay, G., Cropper, M., Brinkman, B.,2001, A\&A, 365, 298L

\bibitem[]{} Pavlov, G.G., Shibanov, Y.A., Zavlin, V.E., 1995, in {\em The Lives of
Neutron Stars}, eds. A.Alpar, U.Kilizoglu \& J. van Paradijs, KLuwer Academic Publishers, p71
\bibitem[]{} Pavlov, G.G., Sanwal, D., Kiziltan, B., G.Garmire, 2001, ApJ, 559, 131

\bibitem[]{}  Petre, R. Becker, C.M., Winkler, P.F. 1996, ApJ, 465, L43

\bibitem[]{} Potekhin, A., Yakovlev, D., 2001, A\&A, 374, 213

\bibitem[]{}  Rajagopal, M., Romani, R.W. 1996, ApJ, 461, 327

\bibitem[]{} Rasio, F.A., Pfahl, E.D., \& Rappaport, S. 2000, ApJ, 532, L47

\bibitem[]{} Rots, A.H., Jahoda, K., Lyne, A.G., 2000, AAS, HEAD meeting \#32, \#33.08

\bibitem[]{} Saito Y., Kawai N., Kamae T., et al., 1997, ApJ, 477, L37

\bibitem[]{} Saito, Y., 1998, Ph.D. thesis, Tokio Univ.~(ISAS Research Note 643)

\bibitem[]{} Sakurai, I., Kawai, N., Torii, K., Negoro, H., Nagase, F., Shibata, S., Becker, W., 2001, PASJ, 53, 535

\bibitem[]{} Sanwal, D., et al., 2002, AAS 200, 7201

\bibitem[]{} Slane, P., Hughes, J.P., Edgar. J., et al., 2001, ApJ, 548, 814

\bibitem[]{}  Stappers, B.W., Gaensler, B.M., Johnston S., 1999, MNRAS, 308, 609

\bibitem[]{}  Swanepoel J.W.H., de Beer C.F., Loots H., 1996, ApJ, 467, 261

\bibitem[]{} Takahashi M., Shibata S., Torii K., Saito Y., Kawai N. 1998, IAU Circ. 7030

\bibitem[]{} Takahashi, M., Shibata, S., Torii, K., Saito, Y., Kawai, N.,
Hirayama, M., Dotani, T., Gunji, S., Sakurai, H., Stairs, I.H., Manchester, R.N., 2001, ApJ, 554, 316 

\bibitem[]{} Tennant, A.F., Becker, W., Juda, M., Elsner, R.F., Kolodziejczak,
 J.J. Murray, S.S., O'Dell, S.L., Paerels, F., Swartz, D.A., Shibazaki, N.,
 Weisskopf, M.C., 2001, ApJ, 554, L173

\bibitem[]{} Torii, K., 1998, Ph.D. thesis, Osaka Univ.~(ISAS Research Note 650)

\bibitem[]{} Torii, K., Kinugasa, K., Toneri, T., Asanuma, T., Tsunemi, H., Dotani, T., 
Mitsuda, K., Gotthelf, E. V., Petre, R., 1998, ApJ, 494L, 207

\bibitem[]{} Torri, K., Gotthelf, E.V., Vasisht, G., Dotani, T., Kinguasa, K., ApJ, 2000, 534, L71

\bibitem[]{}  Toscano, M., Bailes, M., Manchester, R., Sandhu, J.S., 1998, ApJ, 506, 863

\bibitem[]{} Toscano, M., Britton, M.C., Manchester, R., Bailes, M., Sandhu, J.S., Kulkarni, S., Anderson, S.B.,
 1999, ApJ, 523, L171

\bibitem[]{} Tsunemi, H., Miyata, E., Aschenbach, B., Hiraga, J., Akutsu, D., 2000, PASJ, 52, 887

\bibitem[]{}  Tuohy, I.R., Garmire, G.P. 1980, ApJ, 239, L107

\bibitem[]{}  van Straten, W., Bailes, M., Britton, M., Kulkarno, S.R., Anderson, S.B., Manchseter, R.N.,
 Sarkissian, J., 2001, Nat, 412, 158

\bibitem[]{} Verbunt F., Kuiper L., Belloni T., et al, 1996, A\&A, 311, L9

\bibitem[]{}  Wijnands, R., van der Klis, M., 1998, Nat, 394, 344,

\bibitem[]{}  Weisskopf, M.C., Hester, J.J., Tennant, A.F., et al. 2000, ApJ, 536, L81

\bibitem[]{}  Willingale, R., Aschenbach, B., Griffiths, R. G., Sembay, S., Warwick, R. S.,
 Becker, W., Abbey, A. F., Bonnet-Bidaud, J.-M., 2001, A\&A, 365, L212

\bibitem[]{} Zavlin, V.E., Pavlov, G.G., Shibanov, Y.A., 1996, A\&A, 315, 141

\bibitem[]{} Zavlin, V.E., Tr\"umper, J., Pavlov, G.G., 1999, ApJ, 525, 959

\bibitem[]{} Zavlin, V.E., Pavlov, G.G., Sanwal, D., Manchester, R.N., Tr\"umper, J., 
Halpern, J.P., Becker, W., 2002, ApJ, 569, 894

\end{thebibliography}
\end{document}